\documentclass[10pt]{article}
\usepackage{ametsoc2col_noabstract}

\usepackage{comment} 
\usepackage{graphicx}

\usepackage[labelfont={bf,footnotesize},textfont={footnotesize},margin=1.5em]{caption}

\usepackage{microtype}

\usepackage{amsmath,amsfonts,amssymb,bm}

\usepackage{ulem}

\usepackage{hyperref}
\hypersetup{
	hyperindex,
	breaklinks,
	colorlinks=true,
	linkcolor=blue,
	citecolor=purple,
	bookmarks=true,
	bookmarksopen=true,
	bookmarksopenlevel=2,
	pdfstartpage={1},
	pdfstartview={FitH},
	pdfview={FitH 0},
	pdfauthor={B F Farrell and P J Ioannou},
	pdftitle={Statistical State Dynamics: a new perspective on turbulence in shear flow},
 }
\ifthenelse{\boolean{dc}}
{
\usepackage[all]{hypcap}	
}
{}

\usepackage{natbib}
\usepackage{doi}

\usepackage[usenames,dvipsnames]{xcolor}

\def\Tcal{\mathcal{T}}

\newcounter{saveeqn}%

\newcommand{\be}{\begin{equation}}
\newcommand{\ee}{\end{equation}}
\newcommand{\bdm}{\begin{equation*}}
\newcommand{\edm}{\end{equation*}}
\newcommand{\bea}{\begin{eqnarray}}
\newcommand{\eea}{\end{eqnarray}}

\newcommand{\partialf}[2]
{
 \ifthenelse{\equal{#1}{}}{\frac{\partial}{\partial #2}}{\frac{\partial #1}{\partial #2}}
}

\newsavebox{\astrutbox}
\sbox{\astrutbox}{\rule[-5pt]{0pt}{20pt}}

\usepackage{upgreek}

\newcommand{\myabstract}{
{Planetary turbulence is observed to self-organize into large-scale structures such as zonal jets and 
coherent vortices. One of the simplest models that retains the relevant dynamics of turbulent 
self-organization is a barotropic flow in a beta-plane channel with turbulence sustained by random 
stirring. Non-linear integrations of this model show that as the energy input rate of the forcing is 
increased, the homogeneity of the flow is first broken by the emergence of non-zonal, coherent, westward propagating structures and at larger energy input rates by the emergence of zonal jets. The emergence of both non-zonal coherent structures and zonal jets is studied using a statistical theory, Stochastic Structural Stability Theory (S3T). S3T directly models a second-order approximation to the statistical mean turbulent state and allows the identification of statistical turbulent equilibria and study of their stability. Using S3T, the bifurcation properties of the homogeneous state in barotropic beta-plane turbulence are determined. Analytic expressions for the zonal and non-zonal large-scale coherent flows that emerge as a result of structural instability are obtained and the equilibration of the incipient instabilities is studied through numerical integrations of the S3T dynamical system. The dynamics underlying the formation of zonal jets are also investigated. It is shown that zonal jets form from the upgradient momentum fluxes that result from the shearing of the eddies by the emerging infinitesimal large-scale flow. Finally, numerical 
simulations of the nonlinear equations confirm the characteristics (scale, amplitude and phase speed) of the structures predicted by S3T, even in highly non-linear parameter regimes such as the 
regime of zonostrophic turbulence. }}

\begin{document}

\title{\textbf{\large{Emergence of non-zonal coherent structures}}}

\author{\textsc{Nikolaos A. Bakas}\\
\centerline{\textit{\footnotesize{Department of Physics, University of Ioannina, Ioannina, Greece}}}\\
\and
\textsc{Petros J. Ioannou}\thanks{\textit{Corresponding author address:} Petros Ioannou, University of Athens, Department of Physics, Section of Astrophysics, Astronomy and Mechanics, Building IV, Office 32, Panepistimiopolis, 15784 Zografos, Athens, Greece.\newline{E-mail: \href{mailto:pjioannou@phys.uoa.gr}{pjioannou@phys.uoa.gr}}}\\
\textit{\footnotesize{Department of Physics, National and Kapodistrian University of Athens, Athens, Greece}}
}

\ifthenelse{\boolean{dc}}
{
\twocolumn[
\begin{@twocolumnfalse}
\amstitle

\begin{center}
\begin{minipage}{13.0cm}
\begin{abstract}
	\myabstract
	\newline
	\begin{center}
		\rule{38mm}{0.2mm}
	\end{center}
\end{abstract}

\end{minipage}
\end{center}
\end{@twocolumnfalse}
]
}
{
\amstitle
\begin{abstract}
\myabstract
\end{abstract}
\newpage
}



Atmospheric and oceanic turbulence is commonly observed to be organized into spatially and temporally 
coherent structures such as zonal jets and coherent vortices. A simple model that retains the relevant 
dynamics, is a barotropic flow on a $\beta$-plane with turbulence sustained by random stirring. Numerical 
simulations of the stochastically forced barotropic vorticity
equation on the surface of a rotating sphere or on a $\beta$-plane, have shown the coexistence of robust
zonal jets and of large-scale westward propagating coherent structures that are referred to as satellite modes
\citep{Danilov-04} or zonons \citep{Sukoriansky-etal-2008,Galperin-etal-10}. Emergence of
these coherent structures in barotropic turbulence has also another feature. As the energy input of the stochastic
forcing is increased or dissipation is decreased, there is a sudden onset of coherent
zonal flows \citep{Srinivasan-Young-12,Constantinou-etal-2012} and non-zonal coherent structures
\citep{Bakas-Ioannou-2014}. This argues that the emergence of coherent structures in a homogeneous
background of turbulence is a bifurcation phenomenon.

An advantageous method to study such a phenomenon, is to adopt the perspective of statistical state dynamics of
the flow, rather than look into the dynamics of sample realizations of direct numerical simulations. This amounts
to study the dynamics and stability of the statistical equilibria arising in the turbulent flow, which are fixed
points of the equations governing the evolution of the flow statistics. This approach is followed in the Stochastic Structural Stability
Theory (S3T) \citep{FI-03} or Second Order Cumulant Expansion theory (CE2) \citep{Marston-etal-2008}. This theory
is based on two building blocks. The first
is to do a Reynolds decomposition of the dynamical variables into the sum of a mean value that represents the coherent flow
and fluctuations that represent the turbulent eddies and then form the cumulants containing the information on the mean values
(first cumulant) and on the eddy statistics (higher order cumulants). The second building block is to truncate the equations
governing the evolution of the cumulants at second order by either parameterizing the terms involving the third
cumulant \citep{FI-93d,FI-93e,FI-93f,Sole_Farrell-96,DelSole-04} or setting the third cumulant to zero \citep{Marston-etal-2008,Tobias-etal-2011,Srinivasan-Young-12}. Restriction of the dynamics to the first two cumulants is equivalent to
neglecting the eddy-eddy interactions in the fully non-linear dynamics and retaining only the interaction between the eddies
with the instantaneous mean flow. While such a second order closure might seem crude at first sight, there is strong evidence
to support it \citep{Bouchet-etal-2013}.

Previous studies employing S3T have already addressed the bifurcation from a homogeneous turbulent regime to a jet forming
regime in barotropic $\beta$-plane turbulence and identified the emerging jet structures both numerically
\citep{FI-07} and analytically \citep{Bakas-Ioannou-2011,Srinivasan-Young-12} as linearly unstable modes to the homogeneous
turbulent state equilibrium. It was also shown that the resulting dynamical system for the evolution of the
first two cumulants linearized around the homogeneous equilibrium possesses the mathematical
structure of the dynamical system of pattern formation \citep{Parker-Krommes-2013}. Comparison of the results of
the stability analysis with direct numerical simulations have shown that
the structure of zonal flows that emerge in the non-linear simulations can be predicted by S3T
\citep{Srinivasan-Young-12,Constantinou-etal-2012}. However, these research efforts, have assumed that 
the ensemble average employed in S3T is
equivalent to a zonal average, a simplification that treats the non-zonal structures as incoherent and cannot
address their emergence and effect on the jet dynamics. In addition, the eddy-mean flow dynamics underlying
the S3T instability even in the jet formation case, that involve only the interactions of small scale waves with
the large-scale coherent structures are not clear.

So the goals in this article are the following. The first goal is to adopt a more general interpretation of the
ensemble average, in order to address the emergence of coherent non-zonal structures. We adopt the more
general interpretation that the ensemble average is a Reynolds average over the fast turbulent motions
\citep{Bernstein-2009,Bernstein-Farrell-2010}. With this definition of the ensemble mean, we obtain the
statistical dynamics of the interaction of the coarse-grained ensemble average field, which can be zonal or
non-zonal coherent structures represented by their vorticity, with the fine-grained incoherent field
represented by the vorticity second cumulant and we revisit the structural stability of the
homogeneous equilibrium under this assumption. The second goal is to study in detail the eddy-mean flow dynamics
underlying the S3T instability focusing on the example of jet formation. And the third goal is to compare the
characteristics of the structures that emerge in S3T against non-linear simulations, even in highly non-linear
regimes that at first glance present a challenging test for the restricted dynamics of S3T.

\section{Formulation of Stochastic Structural Stability Theory under a generalized average}

Consider a nondivergent barotropic flow on a $\beta$-plane with cartesian coordinates
$\mathbf{x}=(x,y)$. The velocity field, $\mathbf{u}=(u, v)$, is  given by
$(u,v)=(-\partial_y\psi,\partial_x\psi)$, where $\psi$ is the  streamfunction.
Relative vorticity $\zeta(x,y,t)= \Delta \psi$, evolves according to the non-linear (NL) equation:
\begin{equation}
\left ( \partial_t + \mathbf{u}\cdot\nabla \right )\zeta +\beta v=-r\zeta-\nu\Delta^2\zeta+\sqrt{\varepsilon}f^e,\label{eq:derivation1}
\end{equation}
where $\Delta=\partial_{xx}^2+\partial_{yy}^2$ is the horizontal Laplacian, $\beta$ is the gradient of planetary vorticity,
$r$ is the coefficient
of linear dissipation that typically parameterizes Ekman drag in planetary atmospheres and $\nu$ is the
coefficient of hyper-diffusion that dissipates enstrophy flowing into unresolved scales. The exogenous forcing
term $f^e$, parameterizes processes such as small scale convection or baroclinic instability, that are missing
from the barotropic dynamics and is necessary to sustain turbulence. We assume that $f^e$ is a
temporally delta correlated and spatially homogeneous random stirring injecting energy at a rate $\varepsilon$ 
and having a two-point, two-time correlation
function of the form:
\begin{equation}
\left<f^e(x_1,y_1,t_1)f^e(x_2,y_2,t_2)\right>=\delta(t_2-t_1)\Xi(x_1, x_2, y_1, y_2),\label{eq:forc_prop}
\end{equation}
where the brackets denote an ensemble average over the different realizations of the forcing.

S3T describes the statistical dynamics of the first two same time cumulants of (\ref{eq:derivation1}). The 
equations governing the evolution of the first two cumulants are obtained as follows. We decompose the 
vorticity field into  the averaged field, $Z=\Tcal [\zeta]$, defined as a time average over an intermediate 
time scale and deviations from the mean or eddies, $\zeta'=\zeta-Z$. The intermediate time
scale is larger than the time scale of the turbulent motions but smaller than the time scale of the large 
scale motions. With this decomposition we rewrite (\ref{eq:derivation1}) as:
\begin{equation}
\left(\partial_t+\mathbf{U}\cdot\nabla\right)Z+\beta V=-\nabla\cdot\Tcal [\mathbf{u}'\zeta']-rZ-
\nu\Delta^2Z,\label{eq:Q_evo}
\end{equation}
where $\mathbf{U}=[U, V]=[-\partial_y\Psi, \partial_x\Psi]$ and $\mathbf{u}'=[u', v']=
[-\partial_y\psi ', \partial_x\psi ']$ are the mean and the eddy velocity fields respectively.
The mean vorticity is therefore forced by the divergence of the mean vorticity fluxes. The eddy vorticity 
$\zeta'$ evolves according to:
\begin{align}
& \left(\partial_t+\mathbf{U}\cdot\nabla\right)\zeta' +(\beta+\partial_yZ)v'+u'\partial_xZ
=\nonumber\\
 &\qquad=  -r \zeta'
-\nu\Delta^2\zeta'+f^e + \underbrace{\Tcal[ \mathbf{u}'\cdot\nabla \zeta']-
\mathbf{u}'\cdot\nabla \zeta'}_{f^{nl}} ,\label{eq:q_evo}
\end{align}
where $f^{nl}$ is the term involving the non-linear interactions among the turbulent eddies. A closed 
system for the statistical state dynamics is obtained by first neglecting the eddy-eddy term $f^{nl}$ to 
obtain the quasi-linear system,
\begin{align}
&\left(\partial_t+\mathbf{U}\cdot\nabla\right)Z+\beta V =-\nabla\cdot\Tcal [\mathbf{u}'\zeta']-rZ-
\nu\Delta^2Z,\label{eq:Q_evo_eql}\\
& \left(\partial_t+\mathbf{U}\cdot\nabla\right)\zeta' +(\beta+\partial_yZ)v'+u'\partial_xZ
=\nonumber\\
&\hspace{12em}=-r \zeta'-\nu\Delta^2\zeta'+\sqrt{\varepsilon}f^e,\label{eq:q_evo_eql}
\end{align}
In order to obtain the statistical dynamics of the quasi-linear system~\eqref{eq:Q_evo_eql}-\eqref{eq:q_evo_eql} 
we adopt the general interpretation that the ensemble average over the forcing realizations is equal to the 
time average over the intermediate time scale \citep{Bernstein-2009,Bernstein-Farrell-2010}. Under this 
assumption, the slowly varying mean flow $Z$ is also the first cumulant of the vorticity $Z=\left<\zeta\right>$, where 
the brackets denote the ensemble average. The time mean of the vorticity flux is equal 
to the ensemble mean of the flux $\Tcal [\mathbf{u}'\zeta']=\left<\mathbf{u}'\zeta'\right>$. The fluxes 
can be related to the second cumulant $C(\mathbf{x}_1, \mathbf{x}_2, t)\equiv \left<\zeta'(\mathbf{x}_1)
\zeta'(\mathbf{x}_2)\right>$, 
which is the correlation function of the eddy vorticity between the two points $\mathbf{x}_i=(x_i, y_i)$, 
$i=1,2$. We hereafter indicate the dynamic variables that are functions of points $\mathbf{x}_i=(x_i, y_i)$ 
with the subscript $i$, that is $\zeta_i'\equiv\zeta'(\mathbf{x}_i)$. By making the identification that the 
fluxes at point $\mathbf{x}$ are equal to the value of the two variable  function
$\left < \mathbf{u}_1' \zeta_2' \right > $  evaluated at the same point $\mathbf{x}=\mathbf{x}_1=\mathbf{x}_2$,
we write the fluxes as:
\begin{equation}
\left< \mathbf{u}'\zeta'\right>=\left<\mathbf{u}_1'\zeta_2'\right>_{\mathbf{x}_1=\mathbf{x}_2}.
\end{equation}
Expressing the velocities in terms of the vorticity  $[u', v']=[-\partial_y\Delta^{-1}, \partial_x\Delta^{-1}]\zeta'$,
where  $\Delta^{-1}$ is the integral operator that inverts vorticity into the streamfunction field, we obtain
the vorticity fluxes as a function of the second cumulant, in the following manner:
\begin{align}
\left< \mathbf{u}'\zeta'\right>&=\left[\left<u_1'\zeta_2'\right>_{\mathbf{x}_1=\mathbf{x}_2},
\left<v_1'\zeta_2'\right>_{\mathbf{x}_1=\mathbf{x}_2}\right]\nonumber\\ &=\left[-\left<\partial_{y_1}\Delta_1^{-1}
\zeta_1'\zeta_2'\right>_{\mathbf{x}_1=\mathbf{x}_2},\left<\partial_{x_1}\Delta_1^{-1}
\zeta_1'\zeta_2'\right>_{\mathbf{x}_1=\mathbf{x}_2}\right]\nonumber\\ &=
\left[-\left(\partial_{y_1}\Delta_1^{-1}
C\right)_{\mathbf{x}_1=\mathbf{x}_2},\left(\partial_{x_1}\Delta_1^{-1}
C\right)_{\mathbf{x}_1=\mathbf{x}_2}\right].\label{eq:vor_fluxes1}
\end{align}
Consequently, the first cumulant evolves according to:
\begin{align}
& \partial_t Z+UZ_x+V(\beta+Z_y)+rZ+\nu\Delta^2Z=\nonumber\\
&\qquad=\partial_x\left(\partial_{y_1}\Delta_1^{-1}
C\right)_{\mathbf{x}_1=\mathbf{x}_2}-\partial_y\left(\partial_{x_1}\Delta_1^{-1}
C\right)_{\mathbf{x}_1=\mathbf{x}_2}.\label{eq:Q_evo2}
\end{align}

Multiplying (\ref{eq:q_evo2}) for $\partial_t\zeta_1'$ by $\zeta_2'$ and (\ref{eq:q_evo2}) for
$\partial_t\zeta_2'$ by $\zeta_1'$, adding the two equations and taking the ensemble average yields
the equation for the second cumulant $C$:
\begin{equation}
\partial_t C-(A_1+A_2)C=\sqrt{\varepsilon}\left<f_1^e\zeta_2'+f_2^e\zeta_1'\right>,\label{eq:cov_evo1}
\end{equation}
where
\begin{equation}
A_i=-{\bf U}_i\cdot\nabla_i-(\beta+\partial_{y_i}Z)\partial_{x_i}\Delta_i^{-1}+\partial_{x_i}Z
\partial_{y_i}\Delta_i^{-1}-r-\nu\Delta_i^2,\label{eq:op_A}
\end{equation}
governs the dynamics of linear perturbations about the instantaneous mean flow $\mathbf{U}$. The 
right hand side of (\ref{eq:cov_evo1}) is the correlation of the external forcing with vorticity,
which for  delta correlated stochastic forcing is independent of the state of the flow
and is equal at all times to the prescribed forcing covariance:
$\sqrt{\varepsilon}\left<f_1^e\zeta_2'+f_2^e\zeta_1'\right>=\varepsilon\left<f_1^ef_2^e\right>=
\varepsilon\Xi$. Therefore The second cumulant evolves then according to:
\begin{equation}
\partial_t C=(A_1+A_2)C+\varepsilon\Xi  ,\label{eq:cov_evo2}
\end{equation}
and forms with Eq.~(\ref{eq:Q_evo2})  the closed  autonomous system
of  S3T theory  that determines the statistical dynamics of the flow approximated at second order.

The S3T system
has bounded solutions (cf.~Appendix~\hyperref[appA]{A}) and the fixed points $Z^E$ and $C^E$, if they exist, define statistical
equilibria of the coherent structures with vorticity, $Z^E$, in the presence of an eddy field with  second order
cumulant or covariance,
$C^E$. The structural stability of these statistical equilibria
addresses the parameters in the physical system which can lead to abrupt reorganization of the turbulent flow.
Specifically, when an equilibrium of the S3T equations becomes unstable as a physical parameter changes, the
turbulent flow bifurcates to a different attractor. In this work, we show that coherent structures emerge as
unstable modes of the S3T system and equilibrate at finite amplitude. The predictions of S3T regarding the
emergence and characteristics of the coherent structures are then compared to the non-linear simulations of
the stochastically forced barotropic flow.

\section{S3T instability and emergence of finite amplitude large-scale structure}

The homogeneous equilibrium with no mean flow
\begin{equation}
Z^E=0,~~C^E={\Xi\over 2r},\label{eq:equil}
\end{equation}
is a fixed point of the S3T system  (\ref{eq:Q_evo2}) and  (\ref{eq:cov_evo2}) in the absence of hyperdiffusion
(cf.~Appendix~\hyperref[appB]{B}). The linear stability of the
homogeneous equilibrium can be addressed by performing an eigenanalysis of the S3T system linearized about
this equilibrium. The eigenfunctions in this case have the plane wave form
\begin{equation}
\delta Z=Z_{nm}e^{inx+imy}e^{\sigma t}~,~~~
\delta C = C_{nm}(\tilde{x}, \tilde{y})e^{in\overline{x}+im\overline{y}}
e^{\sigma t},
\end{equation}
where $\tilde{x}=x_1-x_2$, $\overline{x}=(x_1+x_2)/2$,
$\tilde{y}=y_1-y_2$, $\overline{y}=(y_1+y_2)/2$, $n$ and $m$ are the $x$ and $y$
wavenumbers of the eigenfunction and $\sigma=\sigma_r+i\sigma_i$ is the eigenvalue with
$\sigma_r=\mathrm{Re} (\sigma)$, $\sigma_i=\mathrm{Im}(\sigma)$ being the growth rate and frequency of the
mode respectively. The eigenvalue $\sigma$  satisfies the non-dimensional equation:
\begin{align}
& \frac{\tilde\varepsilon}{2\pi r^3L_f^2}\int_{-\infty}^{\infty}\int_{-\infty}^{\infty}
d\tilde{k}d\tilde{l}(1-\tilde{N}^2/\tilde{K}^2)\hat{\Xi}(\tilde{k},\tilde{l})\times \nonumber\\
& \qquad\times\frac{(\tilde{m}\tilde{k}-\tilde{n}\tilde{l})
\left[\tilde{n}\tilde{m}(\tilde{k}_+^2-\tilde{l}_+^2)+(\tilde{m}^2-\tilde{n}^2)\tilde{k}_+\tilde{l}_+
\right]}{i\tilde\beta\left(\tilde{k}\tilde{K}_s^2-(\tilde{k}+\tilde{n})\tilde{K}^2\right)+(\tilde\sigma+2)\tilde{K}^2
\tilde{K}_s^2}=\nonumber\\
&\qquad\qquad=(\tilde\sigma+1)
\tilde{N}^2-i\tilde{n}\tilde\beta,\label{eq:dispersion}
\end{align}
where $L_f$ is a characteristic length scale, $\tilde\sigma=\sigma /r$ and $(\tilde n,  \tilde m)= L_f (n,m)$
are the non-dimensional eigenvalue and wavenumbers respectively, $\tilde{\varepsilon}=\varepsilon / ( r^3 L_f^2)$
is the non-dimensional energy injection rate of the forcing, $\tilde{\beta}=\beta L_f /r$
is the non-dimensional planetary vorticity gradient,
\begin{equation}
\hat{\Xi}(k,l)=\frac{1}{2\pi} \int_{-\infty}^{\infty}\int_{-\infty}^{\infty}
\Xi(\tilde{x},\tilde{y})e^{-ik\tilde{x}-il\tilde{y}}
\mathrm{d}\tilde{x}\mathrm{d}\tilde{y}~,\label{eq:eq_stream_cov}
\end{equation}
is the Fourier transform of the forcing covariance, $\tilde{K}^2=\tilde{k}^2+\tilde{l}^2$, $\tilde{K}_s^2=(\tilde{k}+\tilde{n})^2+(\tilde{l}+\tilde{m})^2$,
$\tilde{N}^2=\tilde{n}^2+\tilde{m}^2$, $\tilde{k}_+=\tilde{k}+\tilde{n}/2$ and
$\tilde{l}_+=\tilde{l}+\tilde{m}/2$ (cf.~Appendix~\hyperref[appB]{B}). For a forcing with the mirror symmetry
$\hat\Xi(k,-l)=\hat\Xi(k, l)$ in wavenumber space and for $\tilde n\neq 0$, the eigenvalues satisfy
the relations:
\begin{equation}
\tilde{\sigma}_{(-\tilde{n}, \tilde{m})}=\tilde{\sigma}_{(\tilde{n},
\tilde{m})}^*, \mbox{~and~}\tilde{\sigma}_{(\tilde{n}, -\tilde{m})}=\tilde{\sigma}_{(\tilde{n}, \tilde{m})}, \label{eq:symmet}
\end{equation}
implying  that the
growth rates depend on  $|\tilde{n}|$ and $|\tilde{m}|$. As a result, the plane wave
$\delta Z=\cos(nx+my)$ and an array of localized vortices $\delta Z=\cos (nx)\cos(my)$, have
the same growth rate, despite their different structure. For zonally symmetric perturbations
with  $\tilde n=0$, only the second relation in (\ref{eq:symmet}) holds and (\ref{eq:dispersion}) reduces to
the eigenvalue relation derived by \cite{Srinivasan-Young-12} for the emergence of jets in a
barotropic $\beta$-plane.

\begin{figure}
 \centering
\includegraphics[width=19pc]{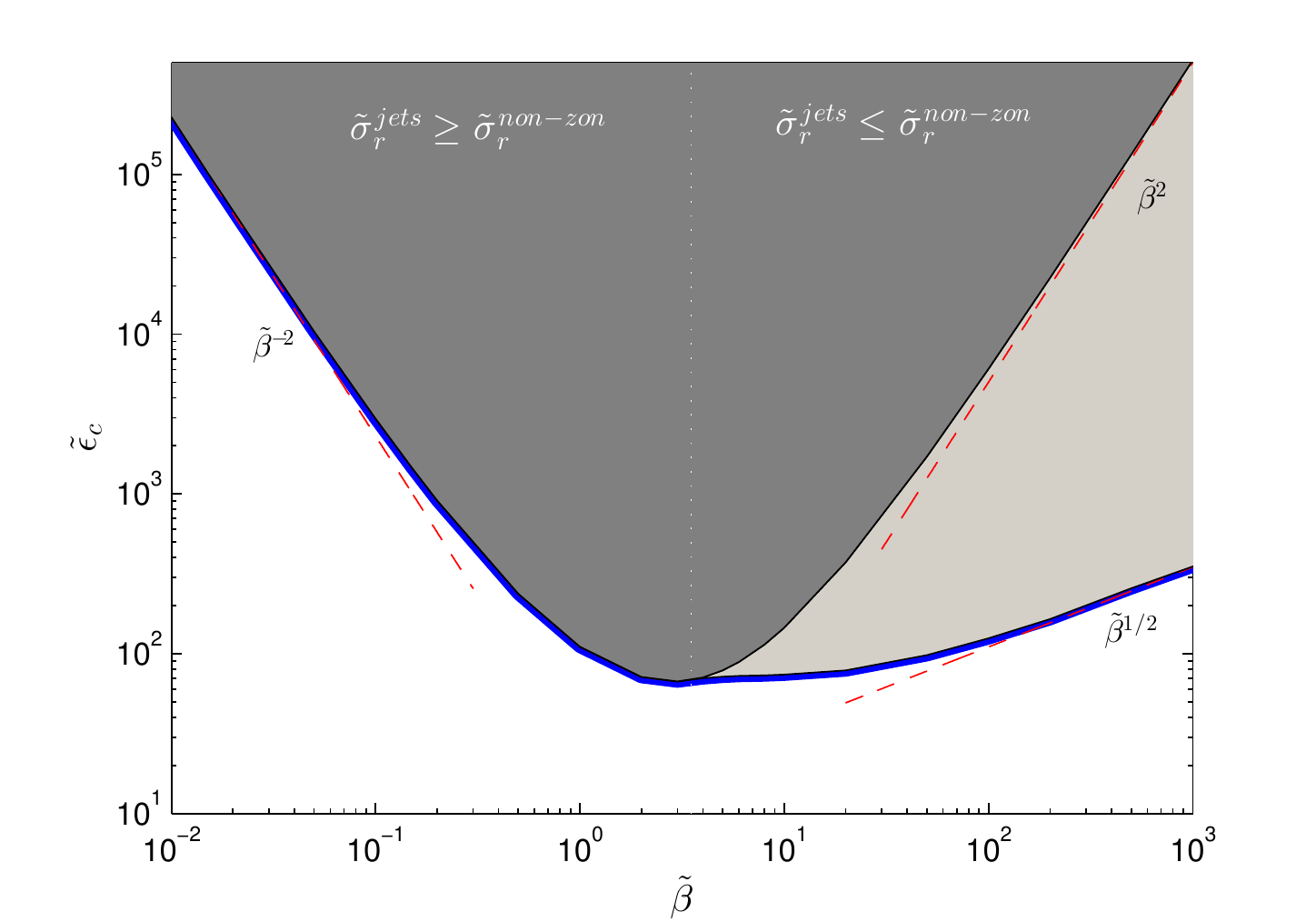}
\caption{The critical energy input rate $\tilde \varepsilon_c$ for structural instability (thick solid line) and
the critical energy input rate for structural instability of zonal jets (solid line) as a function of
$\tilde{\beta}$. The behavior of these critical values for $\tilde \beta\ll 1$ and $\tilde\beta\gg 1$ is  indicated
with the dashed asymptotes. In the light gray region only non-zonal coherent structures emerge, while in the
dark gray region both zonal jets and non-zonal coherent structures emerge. The thin dotted
vertical line $\tilde \beta =\tilde \beta_{min}$ separates the unstable region: for $\tilde \beta<\tilde \beta_{min}$ zonal
structures grow the most, whereas for $\tilde \beta> \tilde \beta_{min}$ non-zonal structures grow the most.}
\label{fig:emin}
\end{figure}

\begin{figure}
 \centering
\includegraphics[width=19pc]{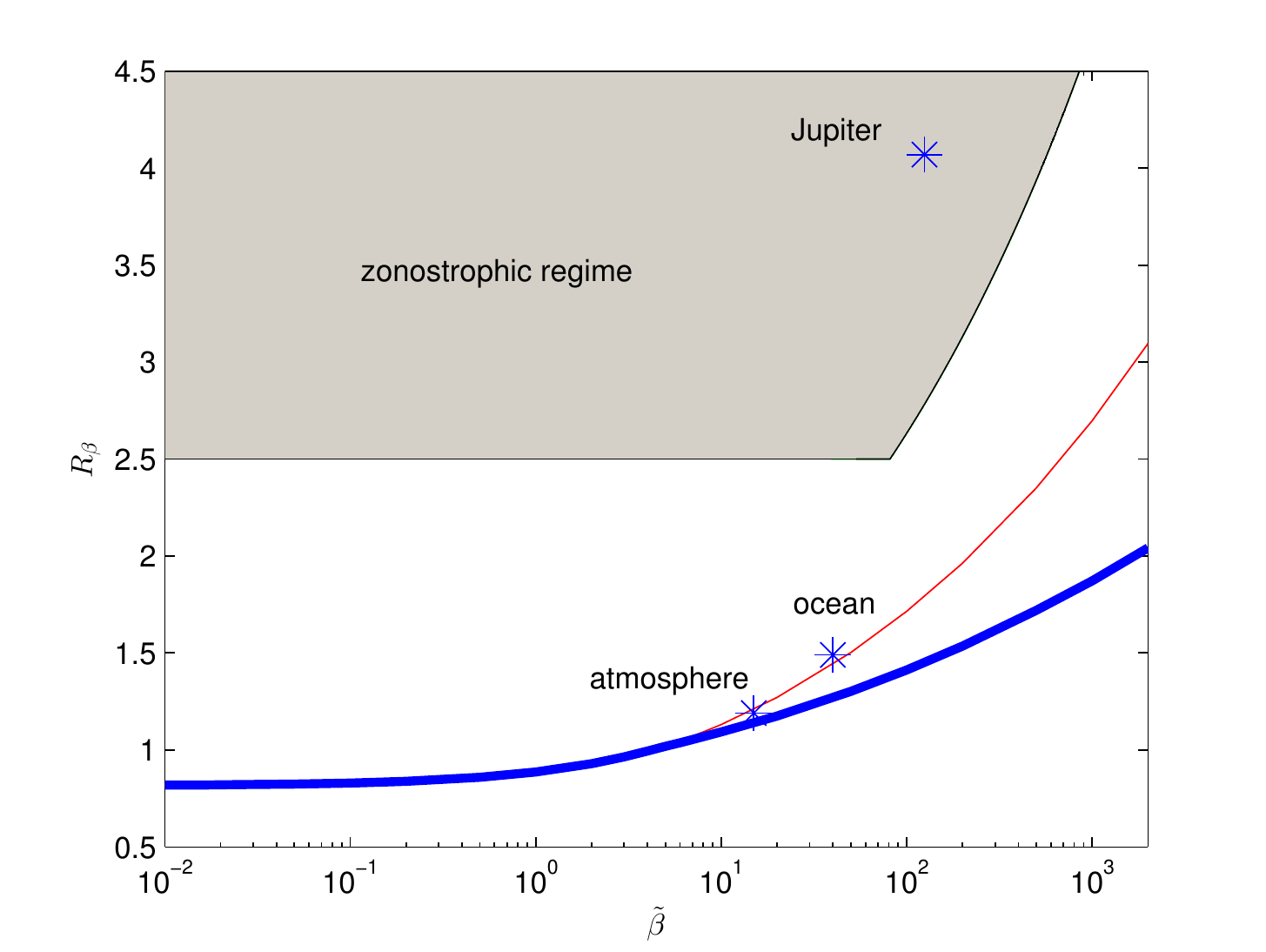}
\caption{The critical zonostrophy parameter $R_\beta=0.7(\tilde\varepsilon_c\tilde\beta^2)^{1/20}$ for
structural instability (thick line) and the corresponding critical parameter for structural instability
of zonal jets (thin line) as a function of $\tilde{\beta}$. The shaded region denotes the zonostrophic
regime for which both the inequalities $R_\beta\geq 2.5$ and $k_\beta/K_f\leq 1/4$ are satisfied. The
stars denote the position of the Earth's atmosphere and ocean as well as the Jovian atmosphere in the
$R_\beta$, $\tilde\beta$ parameter space.}
\label{fig:Rbeta}
\end{figure}

We consider the case of a ring forcing that injects energy at rate $\varepsilon$ at the total wavenumber $K_f$:
\begin{equation}
\hat{\Xi}(k,l)= 2K_f \delta(\sqrt{k^2+l^2}-K_f),\label{eq:iso_for}
\end{equation}
and obtain the eigenvalues $\tilde\sigma$ by numerically solving (\ref{eq:dispersion}).
For small values of the energy
input rate, $\tilde{\sigma}_r<0$ for all wavenumbers and the homogeneous equilibrium
is stable. At a critical $\tilde \varepsilon_c$ the homogeneous flow becomes S3T unstable and exponentially growing
coherent structures emerge. The critical value, $\tilde \varepsilon_c$,  is calculated by first
determining the energy input rate $\tilde{\varepsilon}_t(\tilde{n}, \tilde{m})$
that renders wavenumbers $(\tilde{n}, \tilde{m})$
neutral $\left(\tilde{\sigma}_{r (\tilde{n}, \tilde{m})} =0\right)$, and then by  finding the minimum  energy input rate
over all wavenumbers: $\tilde \varepsilon_c=\mbox{min}_{(\tilde{n}, \tilde{m})} \tilde{\varepsilon}_t$. The critical
energy input rate $\tilde \varepsilon_c$ as a function of $\tilde{\beta}$ is shown in figure~\ref{fig:emin}. In
addition, the corresponding critical zonostrophy parameter $R_\beta=0.7(\tilde\varepsilon_c\tilde\beta^2)^{1/20}$
which was used in previous studies to characterize the emergence and structure of zonal jets in planetary turbulence
\citep{Galperin-etal-10}, is shown as a function of $\tilde\beta$ in figure~\ref{fig:Rbeta}. The absolute
minimum energy input rate required is
$\tilde \varepsilon_c=67$ and occurs at $\tilde{\beta}_{min}=3.5$, while the minimum zonostrophy parameter required for
the emergence of coherent flows is $R_\beta=0.82$ and occurs for $\tilde\beta\rightarrow 0$. For $\tilde{\beta}\leq \tilde{\beta}_{min}$,
the structures that first become marginally stable  are zonal jets (with  $n=0$). The critical input
rate increases as $\tilde \varepsilon_c\sim \tilde{\beta}^{-2}$ for $\tilde{\beta} \rightarrow 0 $ and
the homogeneous equilibrium is structurally stable for all excitation amplitudes when $\tilde \beta =0$.
However, the structural stability for $\tilde \beta=0$ is an artifact of the assumed isotropy of the 
excitation and the assumption of a barotropic flow. In the presence of even the slightest anisotropy 
\citep{Bakas-Ioannou-2011,Bakas-Ioannou-2012}, or in the case of a stratified flow \citep{Parker-Krommes-2014-book}, 
zonal jets are S3T unstable and are expected to emerge even in the absence of $\beta$.
For $\tilde{\beta}>\tilde{\beta}_{min}$, the marginally stable structures are non-zonal and $\tilde \varepsilon_c$ grows as
$\tilde \varepsilon_c\sim \tilde{\beta}^{1/2}$ for $\tilde{\beta} \rightarrow  \infty$. Since the critical forcing for the
emergence of zonal jets (also shown in figure~\ref{fig:emin}), increases as $\tilde \varepsilon_c\sim \tilde{\beta}^2$ for
$\tilde{\beta} \rightarrow  \infty$, for large values of $\tilde{\beta}$ non-zonal structures first emerge and only at
significantly higher $\tilde \varepsilon$ zonal jets are expected to appear. Investigation of these results with other
forcing distributions revealed that the results for $\tilde\beta\gg 1$ are independent of the structure of the
forcing \citep{Bakas-etal-2015}.

The parameter regime of S3T instability is now related to the results of previous studies and to geophysical flows.
Previous studies have identified a parameter regime which is distinguished by robust,
slowly varying zonal jets as well as propagating, non-dispersive, non-zonal coherent structures \citep{Galperin-etal-10}.
This regime that is termed as zonostrophic, is in a region in parameter space in which the zonostrophy parameter is large
($R_\beta\geq 2.5$) and the scale $k_\beta=0.5(\beta^3/\varepsilon)^{1/5}$ in which anisotropization of the turbulent
spectrum occurs is sufficiently larger than the forcing scale ($k_\beta /K_f\leq 1/4$). This regime is shown in figure~\ref{fig:Rbeta}
to be highly supercritical for all $\tilde\beta$. In addition, \cite{Bakas-Ioannou-2014} calculated indicative order of
magnitude values of $\tilde\beta$ and $\tilde\varepsilon$ for the Earth's atmosphere and ocean as well as for the Jovian
atmosphere. From these values we calculated the relevant zonostrophy parameter $R_\beta$ and indicated the
three geophysical flows in figure~\ref{fig:Rbeta}. We can see that all three cases are supercritical: the Jovian atmosphere
is highly supercritical and is well within the zonostrophic regime, while the Earth's atmosphere and ocean are slightly
supercritical (at least within the context of the simplified barotropic model).

\begin{figure}
 \centering
\includegraphics[width=19pc]{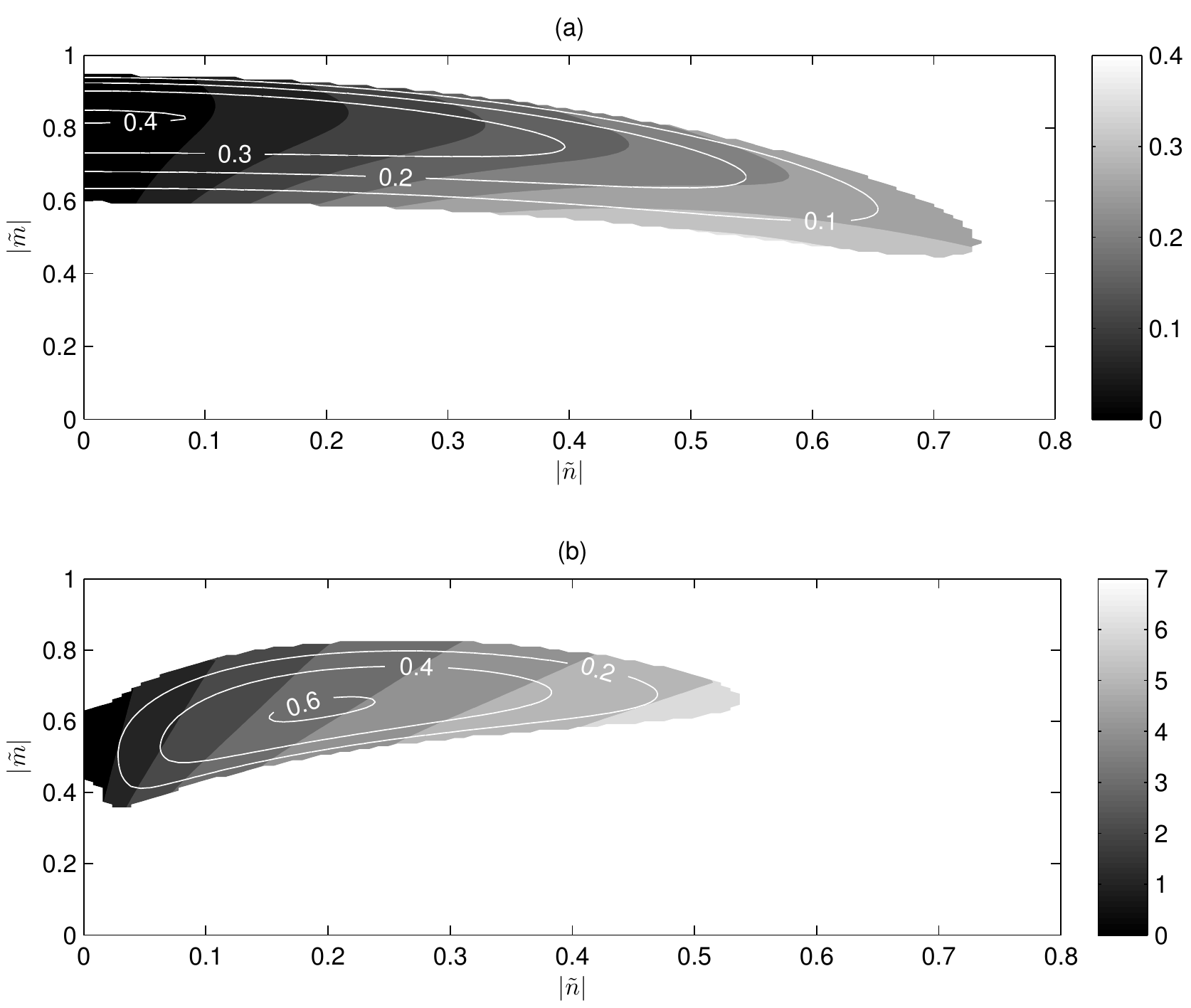}
\caption{Dispersion relation of the unstable modes for $\tilde{\beta}=1$ (panel a) and $\tilde{\beta}=10$ (panel b).
The contours show the growth rate $\tilde{\sigma}_r$ and the shading shows the frequency $\tilde{\sigma}_i$ of the
unstable modes. For $\tilde\beta~O(1)$, stationary zonal jets are more unstable and for $\tilde\beta\gg 1$,
westward propagating non-zonal structures are more unstable. For both panels, the energy input rate is
$\tilde{\varepsilon}=2\tilde \varepsilon_c$.}
\label{fig:growth1}
\end{figure}

We now examine the growth rate and dispersion properties of the unstable modes for
$\tilde{\varepsilon}>\tilde \varepsilon_c$  and consider first the case $\tilde{\beta}=1$,  with
$\tilde \varepsilon=2\tilde \varepsilon_c$. The growth rate of the
maximally growing eigenvalue, $\tilde{\sigma}_r$,  and its associated   frequency of
the mode, $\tilde\sigma_i$, are plotted in figure~\ref{fig:growth1}(a) as a function of $|\tilde{n}|$
and $|\tilde{m}|$. We observe that the region in
wavenumber space defined roughly by $0<|\tilde{n}|<1/2,\mbox{~and~}1/2<|\tilde{m}|<1$ is unstable, with the maximum growth rate
occurring for zonal structures ($\tilde{n}=0$) with $|\tilde{m}|\simeq 0.8$. The frequency of the unstable modes is zero
for zonal jet perturbations ($\tilde n=0$) and non-negative for all other wavenumbers ($\tilde n\ne 0$). Using the symmetries
(\ref{eq:symmet}), this implies that  real unstable  mean flow perturbations $\delta Z$  propagate in the retrograde direction
if $\tilde n \ne 0$ and
are stationary when $\tilde n=0$. As $\tilde \varepsilon$ increases the instability region expands and roughly covers the
sector $1/2<|\tilde{N}|<1$, with zonal structures having a larger growth rate compared to non-zonal
structures, a result that holds for any $\tilde \varepsilon$ when $\tilde \beta < \tilde \beta_{min}$.

\begin{figure}
 \centering
\includegraphics[width=19pc]{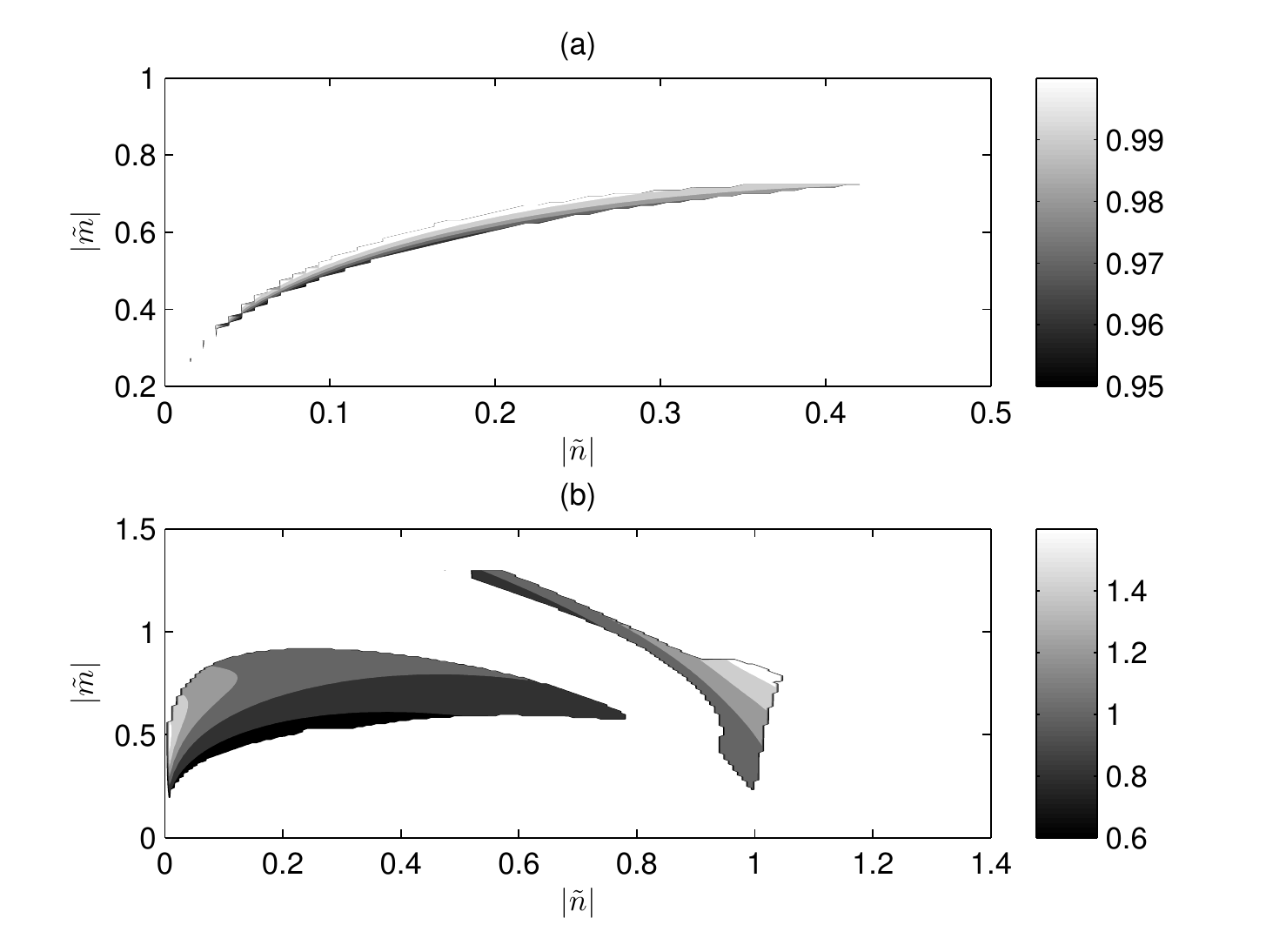}
\caption{Ratio of the frequency of the unstable modes $\tilde{\sigma}_i$ over the
corresponding frequency of a Rossby wave with the same wavenumbers $\tilde\sigma_R$ at (a)
$\tilde{\varepsilon}=2\tilde \varepsilon_c$ and (b) $\tilde{\varepsilon}=50\tilde \varepsilon_c$ when
$\tilde{\beta}=100$. Values of one denote an exact match with the Rossby wave frequency.}
\label{fig:growth_b100_l0}
\end{figure}

For $\tilde \beta > \tilde \beta_{min}$ the non-zonal structures have always larger growth rate. This is illustrated in
figure~\ref{fig:growth1}(b), showing the growth rates and frequencies of the unstable modes for $\tilde{\beta}=10$.
For larger $\tilde \beta$ values there is a tendency for the
frequency of the unstable modes to conform to the corresponding Rossby wave frequency
\begin{equation}
\tilde \sigma_R=\frac{\tilde{\beta}\tilde{n}}{\tilde{n}^2+\tilde{m}^2} ~ ,
\end{equation}
a tendency that does not occur for smaller $\tilde \beta$. A comparison between the frequency of the unstable modes and
the Rossby wave frequency is  shown in figure~\ref{fig:growth_b100_l0} in a plot
of $\tilde{\sigma}_i/\tilde \sigma_R$.  For slightly supercritical
$\tilde \varepsilon$, the ratio is close to one and the unstable modes satisfy the Rossby wave dispersion relation. At higher
supercriticalities though, $\tilde{\sigma}_i$ departs from the Rossby wave frequency (by as much as $40 \%$ for the
case of $\tilde \varepsilon =50 \tilde \varepsilon_c$ shown in figure~\ref{fig:growth_b100_l0}(b)).

\section{Analysis of the eddy-mean flow dynamics underlying jet formation}

In this section, we investigate the eddy-mean flow dynamics leading to jet formation. These dynamics should have
the property of
directly channeling energy from the turbulent motions to the coherent flow without the presence of a
turbulent cascade. Previous studies have identified such mechanisms for the maintenance of zonal jets.
\cite{Huang-98} showed that shear straining of the turbulent field by the jet produced
upgradient momentum fluxes that maintained the jet against dissipation. A simple case that clearly illustrates
the physical picture for the mechanism of shear straining is to consider the evolution of eddies in a planar,
inviscid constant shear flow. The eddies are sheared by the mean flow into thinner elliptical shapes,
while their vorticity is conserved. For an elongated eddy this implies that the eddy velocities decrease
and the eddy energy is transferred to the mean flow through upgradient momentum fluxes. This mechanism can
operate when the time required for the eddies to shear over is much shorter than the dissipation time
scale. The reason is that in this limit even the eddies with streamfunctions leaning against the shear
that initially widen significantly gaining momentum, have the necessary time to shear over, elongate and
surrender their momentum to the mean flow. Given that for an emerging jet the
characteristic shear time scale is necessarily infinitely longer than the dissipation time scale, it
needs to be shown that shear straining can produce upgradient momentum fluxes in this case as well. In
addition, previous studies have shown that shearing of isotropic eddies on an infinite domain does not 
produce any net momentum fluxes \citep{Shepherd-85,Farrell-87,Holloway-10} and should have no effect 
on the S3T instability \citep{Srinivasan-Young-12}. Therefore another mechanism should be responsible 
for producing the upgradient fluxes in the case of an isotropic forcing.

In order to investigate the eddy-mean flow dynamics underlying the S3T instability, we calculate the
vorticity flux divergence that is induced when the statistical equilibrium (\ref{eq:equil}) is perturbed by
an infinitesimal coherent structure $\delta Z$. For an S3T unstable structure, the induced flux divergence
tends to enhance the coherent structure $\delta Z$ producing the positive feedback required for instability.
So the goal of this section is to illuminate the eddy-mean flow dynamics leading to this positive feedback
and to understand qualitatively why the homogeneous equilibrium is more stable for small
and large values of $\tilde\beta$.


For zonal mean flows (\ref{eq:Q_evo2}), (\ref{eq:cov_evo2}) are simplified to:
\begin{equation}
\partial_t U=-\partial_y\left<u'v'\right>-rU=\partial_{y}\left(\partial_{y_2x_1}^2
\Delta_1^{-1}C\right)_{{\bf x}_1={\bf x}_2}-rU,\label{eq:U}
\end{equation}
and
\begin{equation}
\partial_t C=(A_1+A_2)C+\Xi,\label{eq:cov_evo3}
\end{equation}
where
\begin{equation}
A_i=-U_i\partial_{x_i}-(\beta-\partial_{y_iy_i}^2U)\Delta_i^{-1}\partial_{x_i}-r,
\end{equation}
respectively. As a result the zonal mean flow is driven by the momentum flux divergence of
the eddies. The perturbation in vorticity covariance
$\delta C$ that is induced by the mean flow perturbation $\delta U$ can be estimated immediately by assuming
that the system (\ref{eq:U})-(\ref{eq:cov_evo3}) is very close to the stability boundary, so that the growth rate is
small. In this case the mean flow evolves slow enough that it remains in equilibrium with the eddy covariance,
that is $d\delta C/dt\simeq 0$. \cite{Bakas-Ioannou-2012} showed that the ensemble mean momentum flux induced by
an infinitesimal sinusoidal mean flow
perturbation $\delta U=\epsilon\sin(my)$, where $\epsilon\ll 1$ (i.e the eigenfunction of (\ref{eq:Lc})), is
equal in this case to the integral
over time and over all zonal wavenumbers of the responses to all point excitations in the $y$ direction:
\begin{equation}
\delta\left<u'v'\right>=\frac{1}{2\pi}\int_{-\infty}^{\infty}\int_{-\infty}^{\infty}
\int_0^\infty\overline{u'v'}(t)dt d\xi dk,\label{eq:uv_infty}
\end{equation}
where
$\overline{u'v'}(t)$ is the momentum flux at time $t$ produced by:
\begin{equation}
G(k,y-\xi)=B(k)h(y-\xi)e^{ikx+il_0(y-\xi)}.\label{eq:ini_vort}
\end{equation}
The Green's function $G$ has the form of a wavepacket with an amplitude $B(k)$ and a carrier wave with wavenumbers
$(k,l_0)$ that is modulated in the $y$ direction
by the wavepacket envelope $h(y)$. The characteristics of the amplitude, the wavenumber and the envelope depend on the
forcing characteristics, but in any case the calculation of the ensemble mean momentum fluxes is reduced
to calculating the momentum fluxes over the life cycle of wavepackets that are initially at different latitudes
and then adding their relative contributions.

As the wavepacket propagates in the latitudinal direction, its meridional wavenumber and frequency are going to
change due to shearing by the mean flow and due to the change of the mean vorticity gradient $\beta-U_{yy}$. The
resulting time variable momentum flux $\overline{u'v'}(t)$ can be calculated using ray tracing.
According to standard ray tracing arguments, the wave action is conserved along a ray (in the absence of dissipation)
leading to the momentum flux:
\begin{equation}
\overline{u'v'}(t)=-|B|^2A_M(t)e^{-2rt}|h(y-\eta(t))|^2,\label{eq:shear_flux}
\end{equation}
where $A_M(t)=kl_t/(k^2+l_t^2)^2$ is the momentum flux of the carrier wave that
determines the amplitude of the fluxes of the wavepacket and $l_t$, $\eta(t)$ are the time dependent meridional
wavenumber and position of the wavepacket respectively \citep{Andrews-87}.
Because of the small amplitude of the mean flow perturbation $\delta U$, the
wavenumber and position of the packet vary slowly on a time scale $O(\epsilon t)$
compared to the dissipation time scale $1/r$ and the dominant contribution to the time integral
in (\ref{eq:uv_infty}) comes from small times. We can therefore seek asymptotic solutions of the form
\begin{equation}
l_t=l_0+\epsilon l_1+\cdots~ ,~~~~\eta(t)=\xi+c_0t+\epsilon\eta_1(t)+\cdots ~,\label{eq:l_eta_as}
\end{equation}
where $c_0=2\beta kl_0/(k^2+l_0^2)^2$ is the group velocity in the absence of a mean flow and
calculate the integral of $\overline{u'v'}(t)$ over time from the leading order terms.
Substituting (\ref{eq:l_eta_as}) in (\ref{eq:shear_flux}) we obtain:
\begin{align}
\overline{u'v'}(t)&=\underbrace{-|B|^2A_M(0)e^{-2rt}|h(y-\xi-c_0t)|^2}_
{\overline{u'v'}_R} - \nonumber\\
 & \quad \underbrace{-
\epsilon |B|^2\left(\frac{dA_M}{dl_t}\right)_{l_0}l_1(t)e^{-2rt}|h(y-\xi-c_0t)|^2}_
{\overline{u'v'}_S}\nonumber\\
&\quad -\underbrace{\epsilon |B|^2A_M(0)\eta_1(t)e^{-2rt}\frac{d}{dy}|h(y-\xi-c_0t)|^2}_
{\overline{u'v'}_\beta}.\label{eq:shear_flux2}
\end{align}
The first term, $\overline{u'v'}_R$, arises from the momentum flux produced by a wavepacket in the absence
of a mean flow. Because $A_M(0)=kl_0/(k^2+l_0^2)^2$ is odd with respect to wavenumbers, this term does not
contribute to the ensemble averaged momentum flux when integrated over all wavenumbers and will be hereafter
ignored. The second term, $\overline{u'v'}_S$, arises from the
small change in the amplitude of the flux $A_M$ over a dissipation time scale. The third term,
$\overline{u'v'}_{\beta}$, arises from the small change in the position of the packet $\eta$
compared to a propagating packet in the absence of a mean flow. To summarize, the infinitesimal
mean flow refracts the wavepacket due to shearing by the mean flow and due to the change of the 
mean vorticity gradient and slightly changes the amplitude of the fluxes as well as
slightly speeds up or slows down the wavepacket. The sum of these two effects will produce the induced
momentum fluxes.

\subsection{The limit of small scale wavepackets with a short propagation range}

In order to clearly illustrate the behavior of the eddy fluxes, we consider the limit of
$\tilde\beta=\beta L_f/r\ll 1$, where $L_f$ is the scale of the wavepackets and in addition we assume
that the scale of the mean flow, $1/m$, is much larger than the scale of the wavepackets $mL_f\ll 1$.
In this limit, the wavepackets are dissipated before
propagating far from the initial position and the effect of the change in the mean vorticity gradient is
higher order. As a result, \cite{Bakas-Ioannou-2012} show that $l_1$ and $\eta_1$ decrease monotonically
with time with rates independent of $\delta U_{yy}$ and proportional to the shear $\delta U_y(\xi)$
at the initial position $\xi$:
\begin{equation}
l_1=-\delta U_y(\xi) kt~~,~~~~\eta_1=-\beta\delta U_y(\xi)\left(\frac{dA_M}{dl_t}\right)_{l_0}kt^2.
\end{equation}
That is, the amplitude of the flux $A_M$ and the group velocity of the packets
change only due to the shearing of the phase lines of the carrier wave according to the local shear.

\begin{figure}
 \centering
\includegraphics[width=19pc]{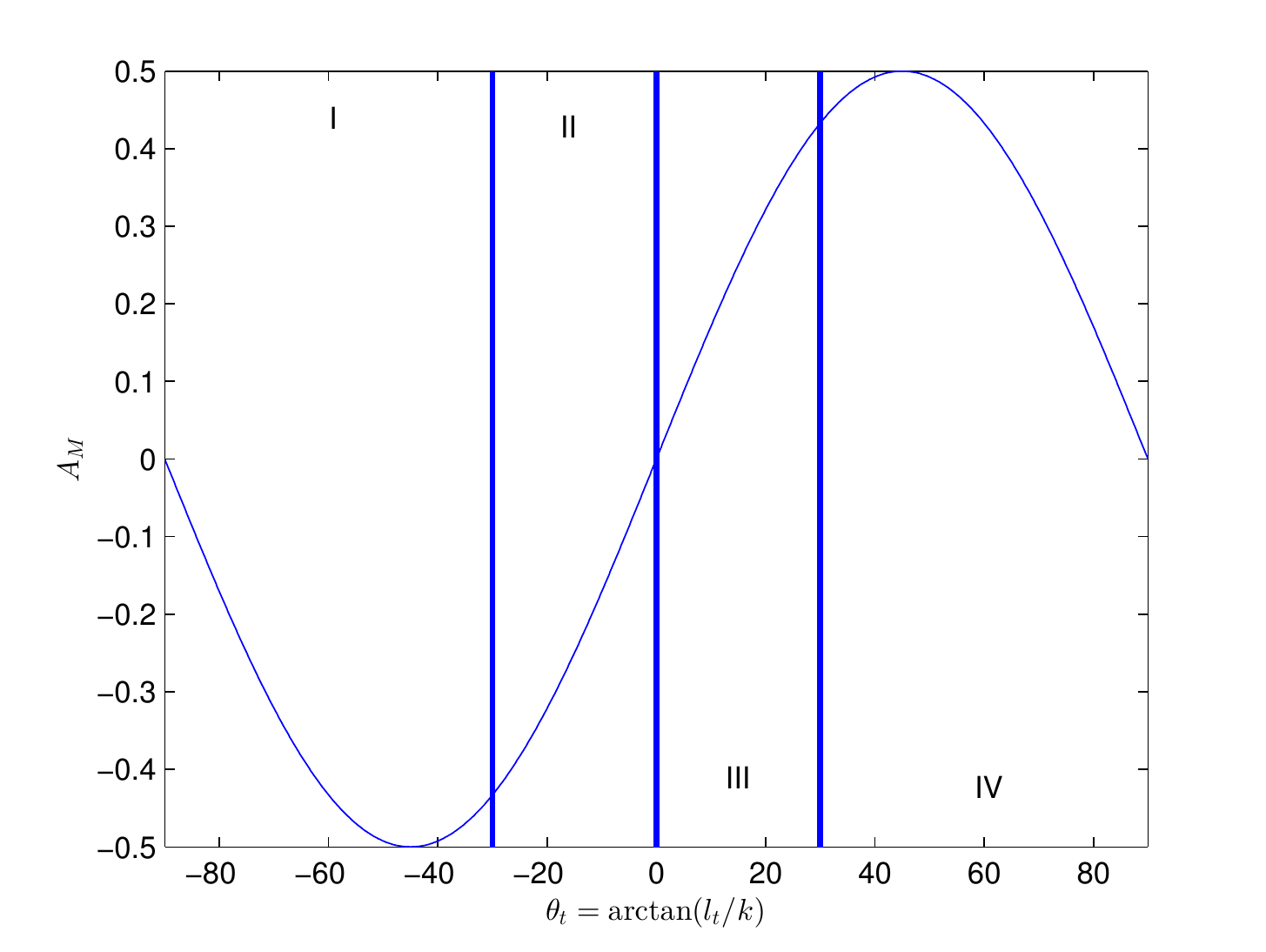}
\caption{Amplitude of the momentum fluxes, $A_M(t)$, of wavepackets as a function of the angle
$\theta_t=\arctan(l_t/k)$ between the phase lines of the central wave and the $y$-axis. The vertical lines separate
the regions with $|\theta_t|<\pi/6$ (II and III) and $|\theta_t|>\pi/6$ (I and IV).}\label{fig:packet1}
\end{figure}

Consider in this limit the first term, $\overline{u'v'}_S$, arising from the small amplitude change. Since the
wavepacket is dissipated before it propagates away, we can ignore to first order propagation:
\begin{align}
\overline{u'v'}_S=&-\epsilon |B|^2\left(\frac{dA_M}{dl_t}\right)_{l_0}l_1(t)e^{-2rt}|h(y-\xi-c_0t)|^2\nonumber\\
\simeq &
|B|^2\delta U_y(\xi)kt\left(\frac{dA_M}{dl_t}\right)_{l_0}e^{-2rt}|h(y-\xi)|^2,\label{eq:uvs}
\end{align}
so that the packet grows/decays in situ. Since the wave packet is rapidly dissipated, the integrated momentum flux
over its life time will be given to a good approximation by the instantaneous change in the flux\footnote{occurring over the dissipation
time scale $1/r$ that is incremental in shear time units} that is proportional to $(dA_M/dl_t)_{l_0}$.
Figure \ref{fig:packet1} illustrates the amplitude of the momentum flux as a function
of the angle $\theta_t=\arctan(l_t/k)$ of the phase lines of the carrier wave of the packet with the $y$-axis. It is
shown that the momentum flux of a wavepacket with $|\theta_0|<\pi/6$ (that is with phase lines close to the
meridional direction) excited in regions II or III, will increase within the dissipation time scale. Compared to an
unsheared wavepacket, this process leads to upgradient momentum flux. The opposite occurs for waves excited in regions
I and IV (with $|\theta_0|>\pi/6$) that produce downgradient flux, as their momentum flux decreases.

\begin{figure}
 \centering
\includegraphics[width=19pc]{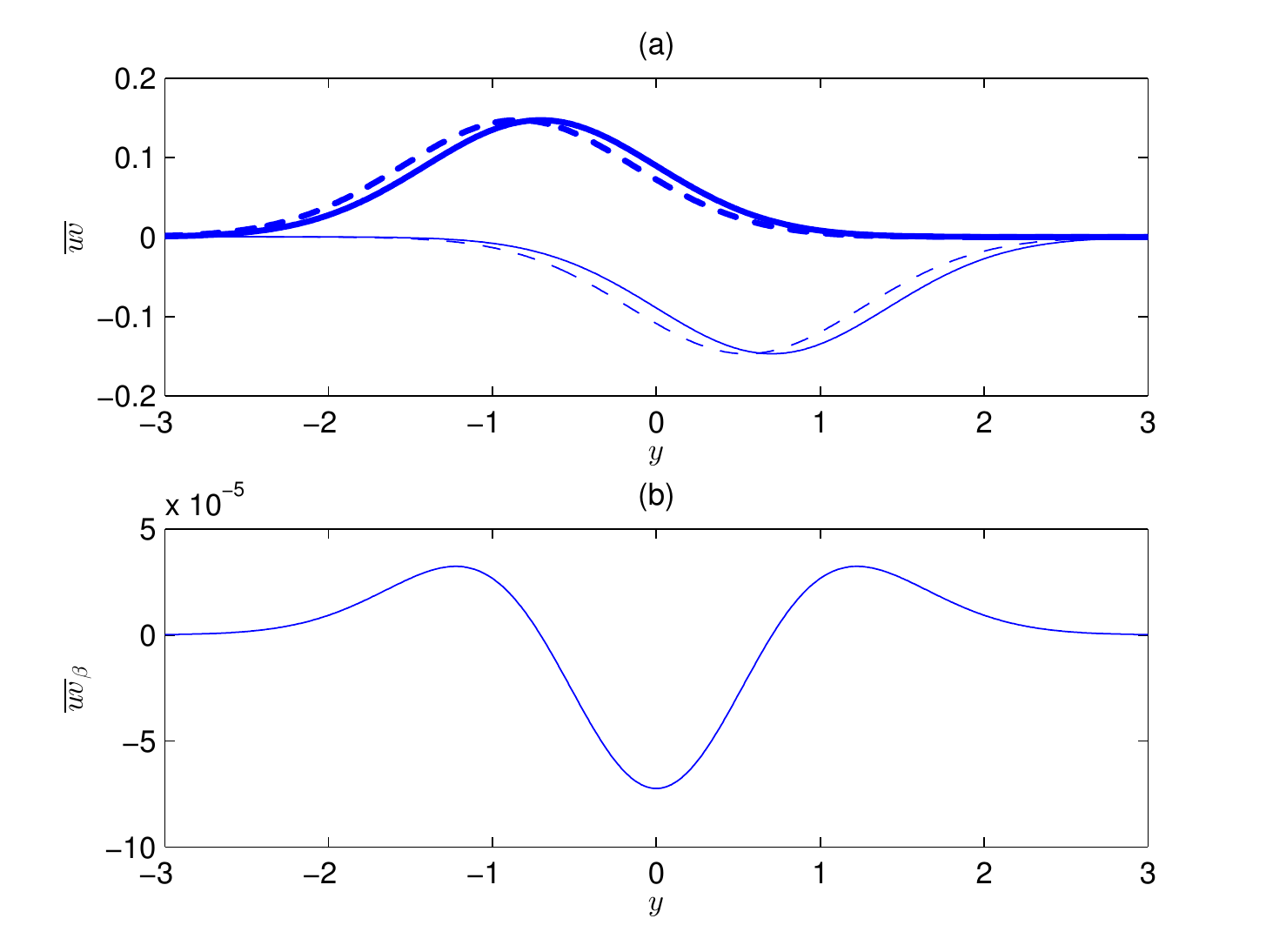}
\caption{(a) Comparison of the momentum fluxes of an unsheared wavepacket excited in
regions II (thick solid line) and III (solid line) to the momentum fluxes of a sheared wavepacket shown by the
corresponding dashed lines, when only the change in propagation is taken into account. A snapshot of the fluxes at
$t=0.2/r$ is shown. The planetary vorticity gradient is $\beta=0.1$, the wavepacket has initial vorticity
$h(y)=e^{-y^2}$, $\sqrt{k^2+l_0^2}=1$, $|\theta_0|=\pi/10$ and $|B|=1$. (b) The difference in momentum
fluxes between a sheared and an unsheared wavepacket calculated over their life cycle.}\label{fig:packet2}
\end{figure}

We now consider the second term, $\overline{u'v'}_\beta$ arising from the effect of propagation on the momentum
flux. The group velocity is given by $c_g=2\beta A_M$ in this case and as a result a wavepacket starting in region
III, will propagate towards the north (c.f. figure~\ref{fig:packet1}). Because shearing slows down the waves in region
III ($\eta_1\sim - (dA_M/dl_t)$), the wavepacket will flux its momentum from southern latitudes compared to when it moved
in the absence of the shear flow. This is shown in figure~\ref{fig:packet2}(a) illustrating the distribution of momentum flux
of an unsheared and a sheared perturbation whose amplitudes are constant. Figure \ref{fig:packet2}(b) plots this difference,
$\overline{u'v'}_\beta$, and shows that the flux is downgradient in this case. The same happens for waves excited in region
II, while the waves excited in regions I and IV produce upgradient flux.

The net momentum fluxes produced by an ensemble of wavepackets, will therefore depend on
the spectral characteristics of the forcing that determine the regions (I-IV), in which the forcing has significant power.
\cite{Bakas-Ioannou-2012} show that for the isotropic forcing (\ref{eq:iso_for}):
\begin{align}
& \delta\left<u'v'\right>=\nonumber\\
&~~={1\over 2\pi}\int_{-\infty}^{\infty}\int_{-\infty}^{\infty}
\overline{u'v'}_S d\xi dk+{1\over 2\pi}\int_{-\infty}^{\infty}\int_{-\infty}^{\infty}
\overline{u'v'}_\beta d\xi dk\nonumber\\
 &~~\simeq 0-{3\tilde\varepsilon\tilde\beta^2r\over 32\pi K_f^4}{d^3\delta U\over dy^3}.\label{eq:integ_step2}
\end{align}
The first integral is zero, because the gain in momentum occurring for $|\theta_0|<\pi/6$
(waves excited in regions II, III) is fully compensated by the loss in momentum for $|\theta_0|>\pi/6$ (waves
excited in regions I, IV) since for the isotropic forcing all possible wave orientations are equally excited.
The net momentum fluxes are therefore produced by the $\overline{u'v'}_{\beta}$ term and are
upgradient, because the loss in momentum occurring for $|\theta_0|<\pi/6$, is over
compensated by the gain in momentum for $|\theta_0|>\pi/6$. The momentum fluxes are also proportional to the
third derivative of $\delta U$ yielding a hyper-diffusive momentum flux divergence that tends to reenforce the mean
flow and is therefore destabilizing. These destabilizing fluxes are proportional to $\tilde\beta^2$ and
as a result, the energy input rate required to form zonal jets increases as $1/\tilde\beta^2$ in this
limit. It is worth noting that the first term integrates to zero only for the special case of
the isotropic forcing, as even the slightest anisotropy yields a non-zero contribution from $\overline{u'v'}_S$.
For example consider the forcing covariance $\Xi(x_1, x_2, y_1, y_2)=\cos\left(k(x_1-x_2)\right)e^{-(y_1-y_2)^2/\delta^2}$
that mimics the forcing of the barotropic flow by the most unstable baroclinic wave, which has zero meridional wavenumber. 
In this case the forcing that is centered at $l_0=0$ in wavenumber space, injects significant power 
in a band of waves in regions II and III and therefore $\overline{u'v'}_S$ yields upgradient fluxes.

\subsection{The effects of the change in the mean vorticity gradient and the finite propagation range}

In order to take into account the effect of the change in the vorticity gradient, we retain higher order
terms with respect to $mL_f\ll 1$ in $l_1$ and $\eta_1$. In this case it can be shown that $l_1$ decreases
with time at a rate proportional to $U_y(\xi)+U_{yyy}(\xi)$ \citep{Bakas-Ioannou-2012}. Since the local shear
and the local change in the vorticity gradient have different signs, the wavepacket is 'sheared less' and
as a result we expect reduced momentum fluxes compared to the limit discussed in section 4.2. Indeed, for
the isotropic forcing:
\begin{equation}
\delta\left<u'v'\right>\simeq -{3\tilde\varepsilon\tilde\beta^2r\over 32\pi K_f^4}\left({d^3\delta U\over dy^3}-
\frac{1}{4K_f^2}\frac{d^5\delta U}{dy^5}\right).\label{eq:iso_neffect}
\end{equation}
That is, the change in the mean vorticity gradient has a stabilizing effect.

We finally relax the assumption that $\tilde\beta\ll 1$. In this case, $l_1$ and $\eta_1$
are affected by an integral shear and mean vorticity gradient over the region of propagation.
For larger $\tilde\beta$, the wavepacket will encounter regions of both positive and negative shear
and as a result, the momentum fluxes that are qualitatively proportional to the integral shear
over the propagation region will be reduced. In the limit $\tilde\beta\gg 1$, the
region of propagation is the whole sinusoidal flow with consecutive regions
of positive and negative shear and the integral shear along with the fluxes will asymptotically
tend to zero. As a result, the energy input rate required for structural instability of zonal jets
increases with $\tilde\beta$ in this limit.

\section{Equilibration of the S3T instabilities}


We now investigate the equilibration of the instabilities by studying the S3T system
(\ref{eq:Q_evo2}), (\ref{eq:cov_evo2}) discretized in a doubly periodic channel of size
$2\pi \times 2 \pi$. We approximate the monochromatic forcing (\ref{eq:iso_for}), by
considering the narrow band forcing
\begin{equation}
\hat{\Xi}(k, l)=\frac{K_f}{\Delta K_f}\left\{\begin{array}{ll} 1,~\mbox{for~}|\sqrt{k^2+l^2}-K_f|\leq \Delta
K_f\\0,~\mbox{for~}|\sqrt{k^2+l^2}-K_f|>\Delta K_f\end{array}\right. , \label{eq:finite_ring}
\end{equation}
where $k$, $l$ assume integer values, that injects energy at rate $\varepsilon$ in a narrow ring
in wavenumber space with radius $K_f$ and width $\Delta K_f$. We consider the set of parameter values
$\beta=10$, $r=0.01$, $\nu=1.19\cdot 10^{-6}$, $K_f=10$ and $\Delta K_f=1$, for which $\tilde{\beta}=100$.
The integration is therefore in the parameter region of figure~\ref{fig:emin} in which  the  non-zonal
structures are  more unstable than the zonal jets. The growth rates of the coherent structures for
integer values of the wavenumbers,  $n$ and $m$  are  calculated from the discrete version of equation
(\ref{eq:dispersion}) obtained by substituting the integrals with sums over integer values of the
wavenumbers \citep{Bakas-Ioannou-2013}.

\begin{figure}
 \centering
\includegraphics[width=19pc]{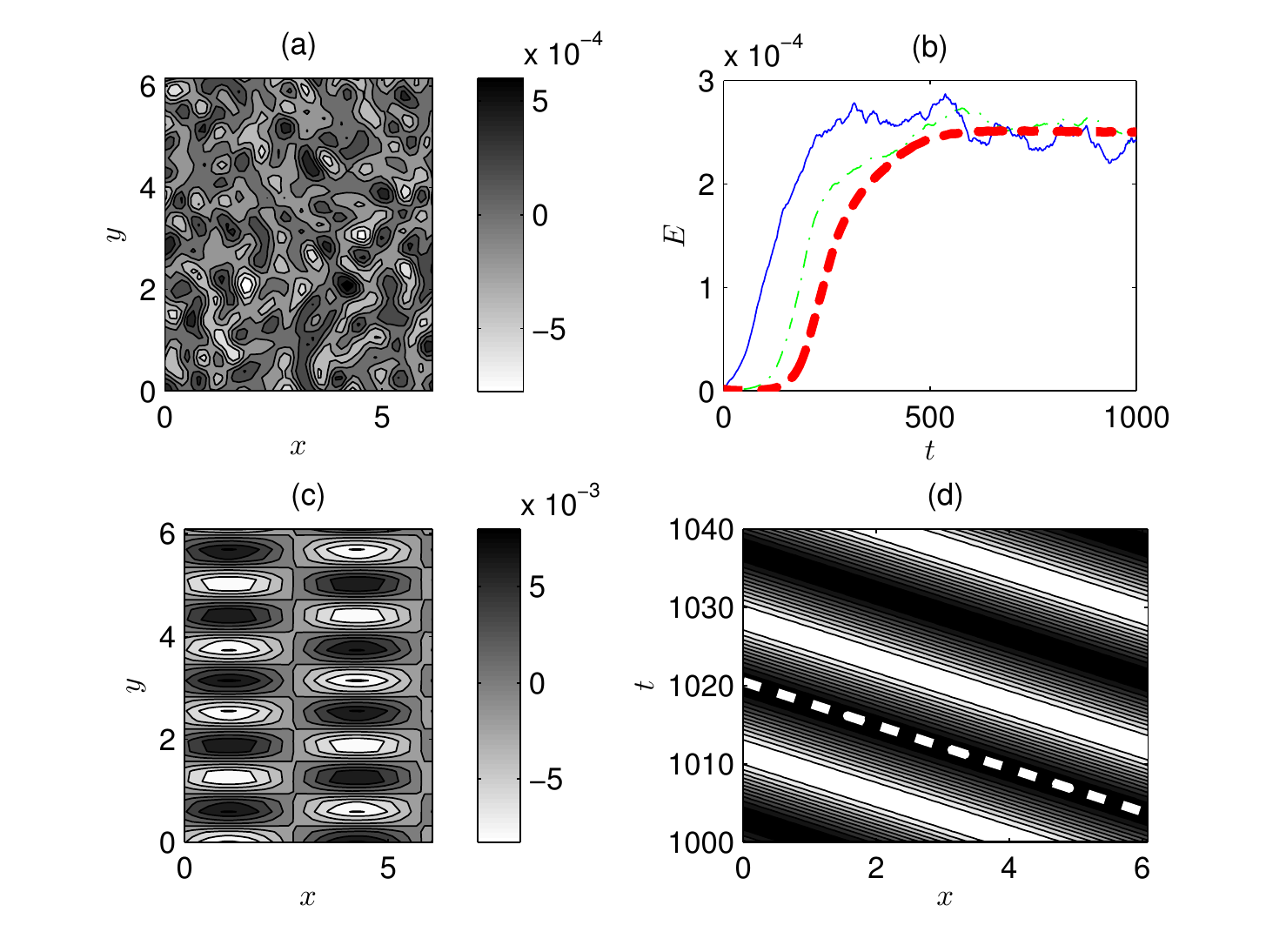}
\caption{Equilibration of the S3T instabilities. (a) Streamfunction of the initial perturbation. (b) Energy
evolution of the initial perturbation shown in panel (a) as obtained from the integration of the S3T equations (\ref{eq:Q_evo2})
and (\ref{eq:cov_evo2}) (dashed line) and from the integration of the ensemble quasi-linear (EQL)
system (\ref{eq:q_evo})-(\ref{eq:Q_evo}) with $N_{\textrm{ens}}=10$ (solid line) and $N_{\textrm{ens}}=100$ (dash-dotted line) ensemble members
that is discussed in section 6. (c) Snapshot of the streamfunction $\Psi_{eq}$
of the traveling wave structure and (d) Hovm\"oller diagram of $\Psi_{eq}(x,y=\pi/4,t)$ for the finite equilibrated traveling wave.
The thick dashed line shows the phase speed obtained from the stability equation (\ref{eq:dispersion}). The energy input rate
is $\tilde{\varepsilon}=4\tilde{\varepsilon}_c$ and $\tilde{\beta}=100$.}
\label{fig:equil1}
\end{figure}

We first consider the supercritical energy input rate
$\tilde \varepsilon=4\tilde \varepsilon_c$. For these parameters only non-zonal modes are unstable, with
the perturbation with $(n, m)=(1, 5)$ growing the most. At $t=0$, we introduce a small random
perturbation, whose streamfunction is shown in figure~\ref{fig:equil1}(a). After a few e-folding times,
a harmonic structure of the form $Z=\cos(x)\cos(5y)$ dominates the large-scale flow. The energy of this large
scale structure shown in figure~\ref{fig:equil1}(b), increases rapidly and eventually saturates. At this point
the large-scale flow gets attracted to a traveling wave finite amplitude equilibrium structure
(cf.~figure~\ref{fig:equil1}(c)) close in form to the harmonic $Z=\cos(x)\cos(5y)$ that propagates westward.
This is illustrated in the Hovm\"oller diagram of $\psi(x, y=\pi/4, t)$ shown in \ref{fig:equil1}(d). The
sloping dashed line in the diagram corresponds to the phase speed of the traveling wave, which is found to
be approximately the phase speed of the unstable $(n, m)=(1, 5)$ eigenmode.

\begin{figure}
 \centering
\includegraphics[width=19pc]{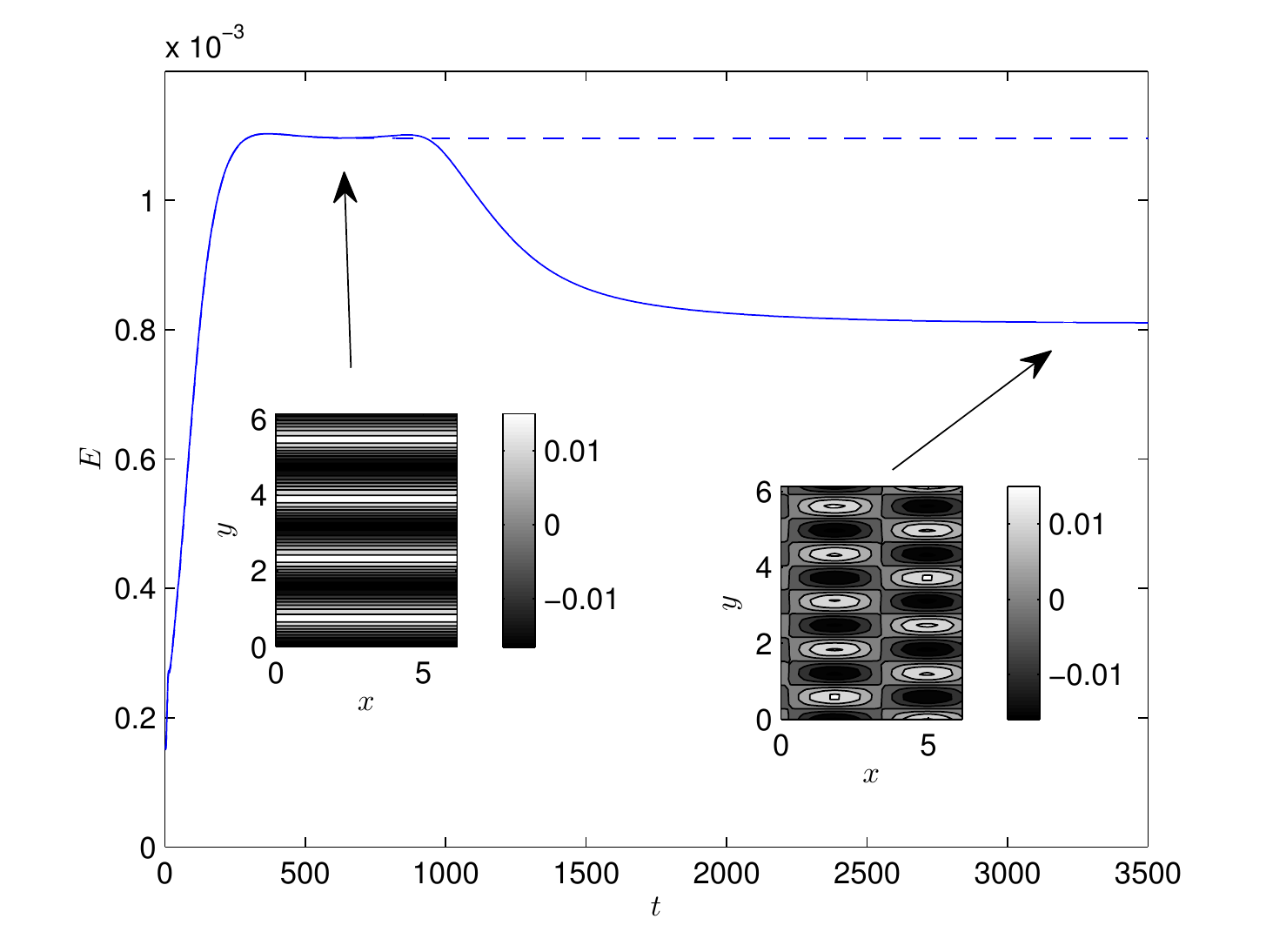}
\caption{Energy evolution of an initial jet perturbation $\delta Z=0.1\cos(4y)$ for the zonally restricted S3T dynamics
(\ref{eq:U})-(\ref{eq:cov_evo3}) (dashed line) and the generalized S3T dynamics
(\ref{eq:Q_evo2}), (\ref{eq:cov_evo2}) (thin line). The insets show a snapshot of the mean flow streamfunction at
$t=700$ (left) and the streamfunction of the equilibrated structure at $t=3500$ (right) under the generalized S3T dynamics.
The parameters are $\tilde{\varepsilon}=10\tilde\varepsilon_c$ and $\tilde{\beta}=100$.}
\label{fig:equil2}
\end{figure}

Consider now the energy input rate $\tilde \varepsilon=10\tilde \varepsilon_c$. While the maximum growth rate still
occurs for the $(|n|, |m|)= (1, 5)$ non-zonal structure, zonal jet perturbations are unstable as well. If the S3T dynamics
are restricted to account only for the interaction
between zonal flows and turbulence by employing a zonal mean rather than an ensemble mean, an infinitesimal jet perturbation
will grow and equilibrate at finite amplitude. To
illustrate this we integrate the S3T dynamical system (\ref{eq:U})-(\ref{eq:cov_evo3}) restricted to zonal flow coherent
structures. The energy of the small zonal jet perturbation $\delta Z=0.1\cos(4y)$ is shown in figure~\ref{fig:equil2} to
grow and saturate at a constant value and the streamfunction of the equilibrated jet is shown in the left inset in figure
\ref{fig:equil2}. However, in the context of the generalized S3T analysis that takes into account the dynamics of the
interaction between coherent non-zonal structures and jets, we find that these S3T jet equilibria can be saddles: stable to
zonal jet perturbations but unstable to non-zonal perturbations. To show this, we consider the evolution of the same jet
perturbation $\delta Z=0.1\cos(4y)$ under the generalized S3T dynamics (\ref{eq:Q_evo2}), (\ref{eq:cov_evo2}) and find
that the flow follows the zonally restricted S3T dynamics and equilibrates to the same finite amplitude zonal jet (cf.
figure~\ref{fig:equil2}). At this point we insert a small random perturbation to the equilibrated flow. Soon after, non-zonal
undulations grow and the flow transitions to the stable $Z=\cos(x)\cos(5y)$ traveling wave state that is also shown in
figure~\ref{fig:equil2}. As a result, the finite equilibrium zonal jet structure is S3T unstable to coherent non-zonal
perturbations and is not expected to appear in non-linear simulations despite the fact that the zero flow equilibrium is
unstable to zonal jet perturbations. We will elaborate more on this issue in the next section.

\begin{figure}
 \centering
\includegraphics[width=19pc]{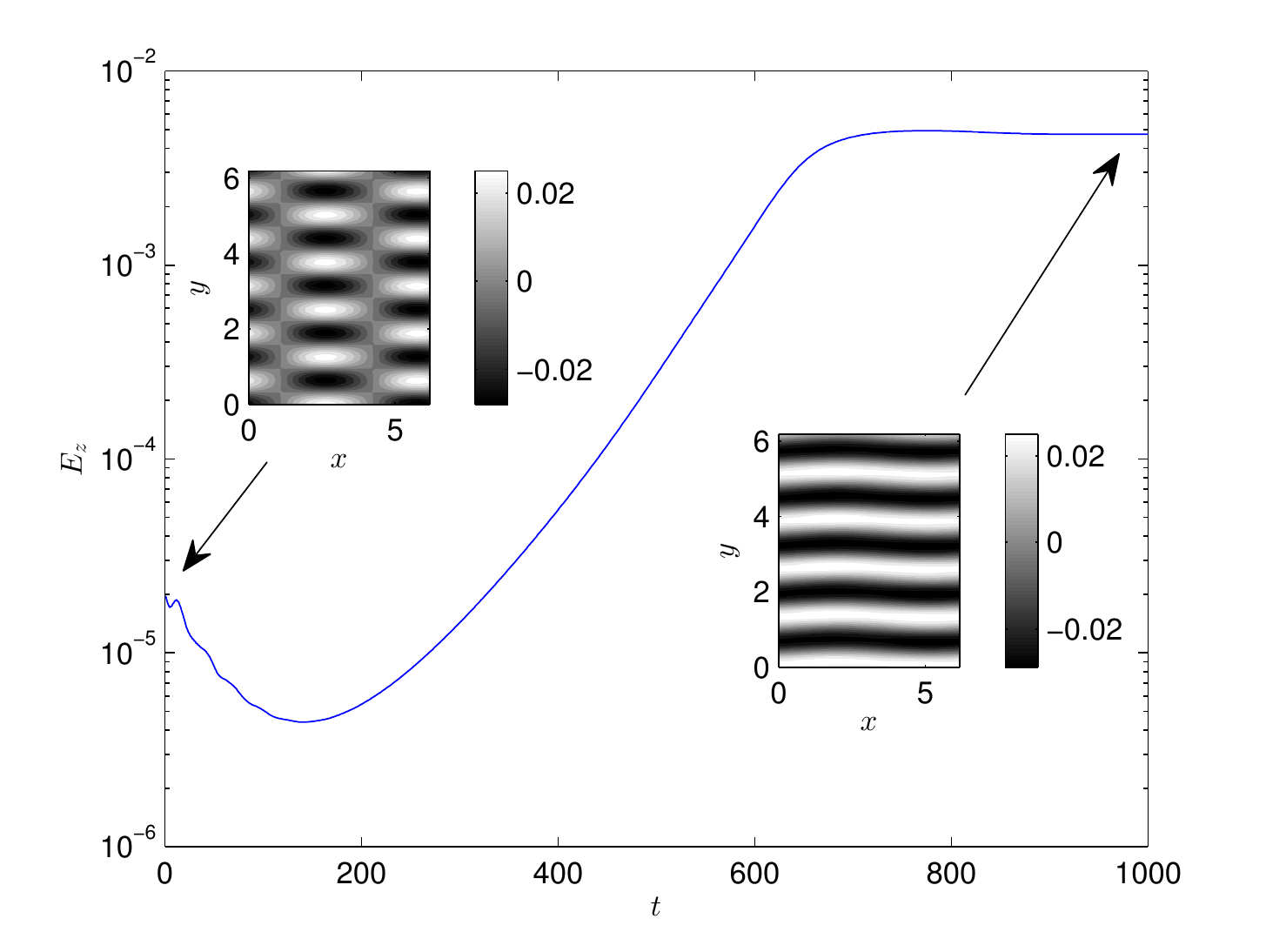}
\caption{Zonal energy evolution of a random zonal perturbation imposed on the non-zonal traveling wave
equilibrium shown in the left inset. The streamfunction of the equilibrated structure is shown in the right inset.
The energy input rate is $\tilde{\varepsilon}=30\tilde\varepsilon_c$ and $\tilde{\beta}=100$.}
\label{fig:equil3}
\end{figure}

\begin{figure}
 \centering
\includegraphics[width=19pc]{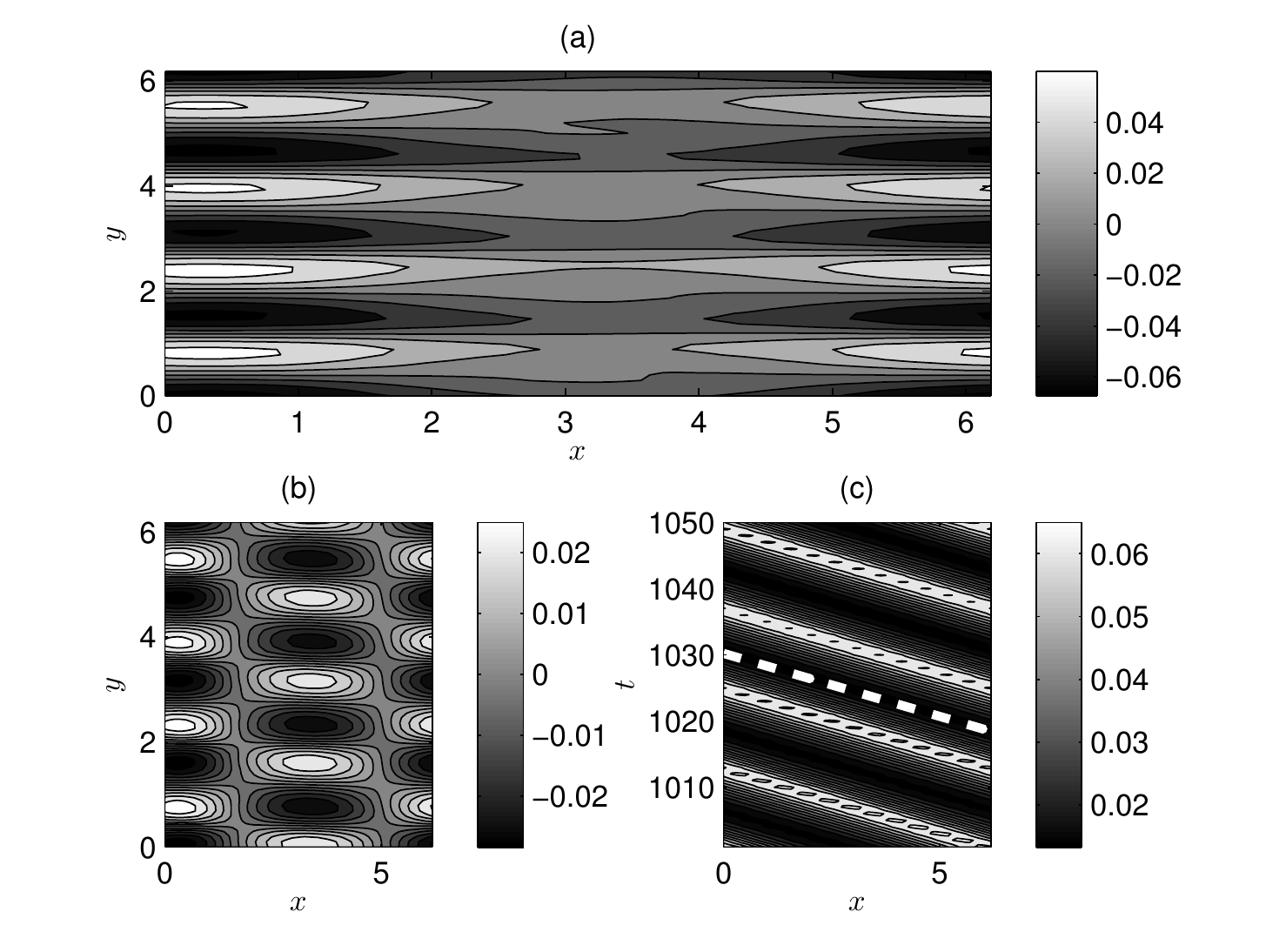}
\caption{Mixed zonal jet-traveling wave S3T equilibrium for $\tilde{\varepsilon}=50\tilde\varepsilon_c$ and $\tilde{\beta}=100$.
(a) Snapshot of the streamfunction $\Psi_{eq}$ of the equilibrium state. (b) Contour plot of the non-zonal component
$\Psi_{eq}-\overline{\Psi_{eq}}$ of the equilibrium structure, where the overline denotes a zonal average. (c)
Hovm\"oller diagram of $\Psi_{eq}(x,y=\pi/4,t)$ for the equilibrated structure.}
\label{fig:equil4}
\end{figure}

Finally, consider the case $\tilde\varepsilon=30\tilde\varepsilon_c$. At this energy input rate, the finite amplitude non-zonal
traveling wave equilibria become S3T unstable. To show this, we consider the non-zonal traveling wave equilibrium
obtained by the evolution of the small non-zonal perturbation $\delta Z=0.01\cos(x)\cos(5y)$ to the homogeneous state
that is shown at the left inset in figure~\ref{fig:equil3} and impose a small random zonal perturbation. The evolution
of the zonal energy $E_z=(1/2)\overline{U}^2$, where the overbar denotes a zonal average, is shown in figure
\ref{fig:equil3}. After an initial transition period, the zonal perturbations grow exponentially and the flow transitions
to the jet equilibrium state shown at the right inset in figure~\ref{fig:equil3}. Note however, that the jet equilibrium
structure is not zonally symmetric. This is a new type of S3T equilibrium: it is a mix between a zonal jet and a non-zonal
traveling wave with the same meridional scale. These mixed equilibria appear to be the attractors for larger energy input
rates as well. This is illustrated in figure~\ref{fig:equil4} showing the structure of the mixed equilibrium at
$\varepsilon=50\varepsilon_c$. The equilibrium structure consists of a large amplitude zonally symmetric jet with larger
scale compared to the mixed state in figure~\ref{fig:equil3}. Embedded in it are non-zonal
vortices with the same meridional scale and with about $14\%$ the energy of the zonal jet. These vortices that are shown in
figure~\ref{fig:equil4}(b) to
have approximately the compact support structure $\Psi=\cos(x)\cos(4y)$ propagate westward as shown in the Hovm\"oller
diagram in figure~\ref{fig:equil4}(c).

\section{Comparison to ensemble mean quasi-linear and non-linear simulations}

\subsection{Comparison to an ensemble of quasi-linear simulations}

Within the context of the second order cumulant closure, the S3T formulation allows
the identification of statistical turbulent equilibria in the infinite ensemble limit,
in which the fluctuations induced by the stochastic forcing are averaged to zero. However,
these S3T equilibria and their
stability properties are manifest even in single realizations of the turbulent system.
For example, previous studies using S3T obtained zonal jet equilibria in barotropic, shallow water
and baroclinic flows in close correspondence with observed jets in planetary flows \citep{FI-07,FI-08,Farrell-Ioannou-2009-closure,Farrell-Ioannou-2009-equatorial}. In
addition, previous studies of S3T dynamics restricted to the interaction between zonal flows and
turbulence in a $\beta$-plane channel showed that when the energy input rate is such that the zero
mean flow equilibrium is unstable, zonal jets also appear in the non-linear simulations with the
structure (scale and amplitude) predicted by S3T \citep{Srinivasan-Young-12,Constantinou-etal-2012}.

A very useful intermediate model that retains the wave-mean flow dynamics of the S3T system while
relaxing the infinite ensemble approximation is the quasi-linear system 
(\ref{eq:Q_evo_eql})-(\ref{eq:q_evo_eql}). Under the ergodic assumption, this can be interpreted 
as an ensemble of quasi-linear equations (EQL) in which the ensemble mean can be calculated from a 
finite number of ensemble members. Its integration is done as follows. A pseudo-spectral code with a 
$128\times 128$ resolution and a fourth order Runge-Kutta scheme for time stepping is used to integrate 
(\ref{eq:Q_evo_eql})-(\ref{eq:q_evo_eql}) forward. At each time
step, $N_{\textrm{ens}}$ separate integrations of (\ref{eq:q_evo_eql}) are performed with the eddies evolving
according to the instantaneous flow. Then the ensemble average vorticity flux divergence is calculated as the
average over the $N_{\textrm{ens}}$ simulations and (\ref{eq:Q_evo_eql}) is stepped forward in time according to those fluxes.
The EQL system reaches a statistical equilibrium at time scales of the order of $t_{eq}\sim O(1/r)$ and the
integration was carried on until $t=100t_{eq}$ in order to collect accurate statistics.

\begin{figure}
 \centering
\includegraphics[width=19pc]{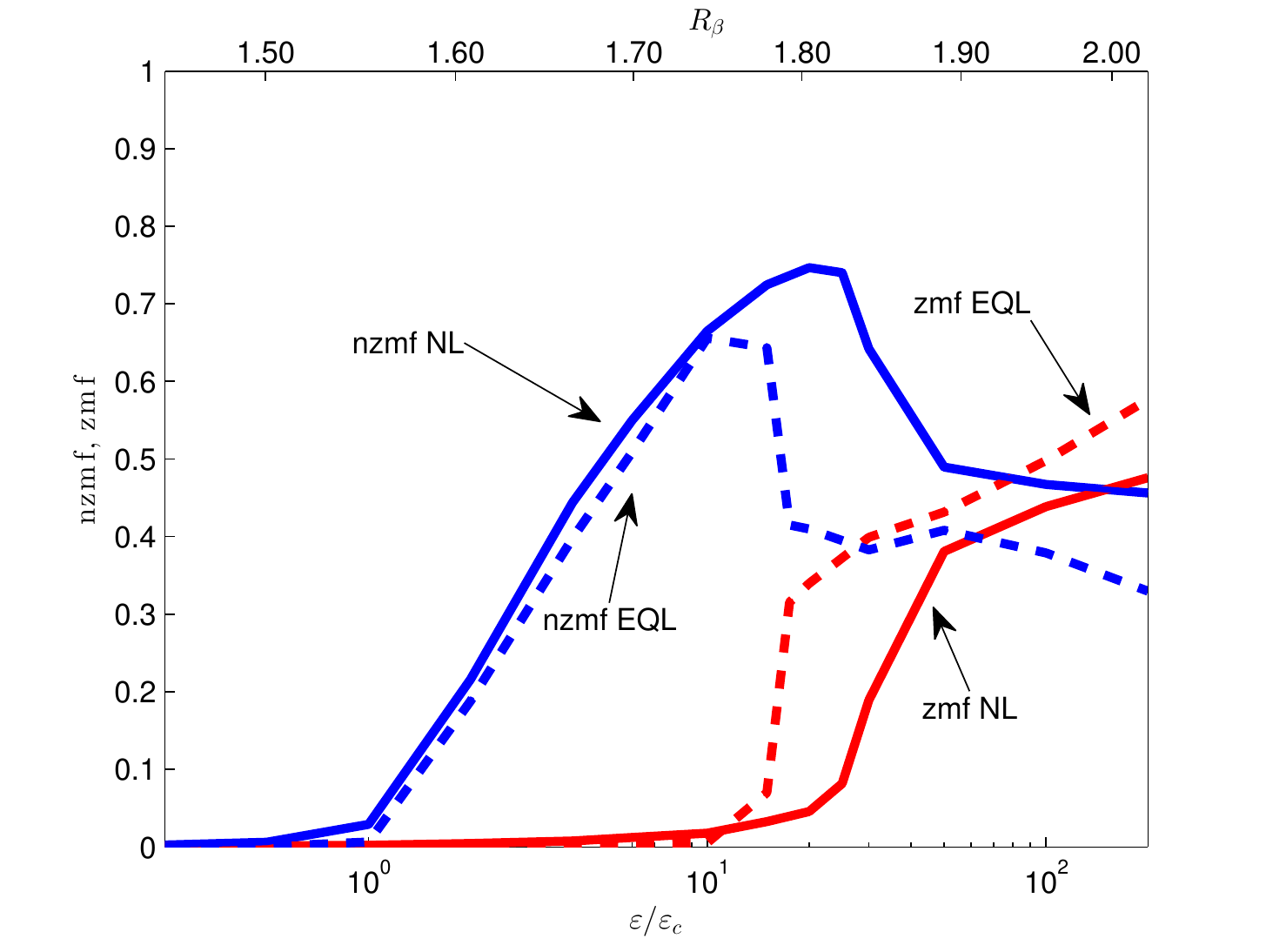}
\caption{The \mbox{zmf} and \mbox{nzmf} indices defined in (\ref{eq:zmf}) and
(\ref{eq:nzmf}) respectively, as a function of energy input rate $\varepsilon/\varepsilon_c$ and the
zonostrophy parameter $R_\beta$ for the non-linear
(NL) integrations and an ensemble of quasi-linear (EQL) integrations (dashed line) with $N_{\textrm{ens}}=10$ ensemble members
as described in section 6. The critical value $\varepsilon_c=8.4\cdot 10^{-6}$ is the energy input rate at which the S3T predicts
structural instability of the homogeneous turbulent state. Zonal jets emerge for $\varepsilon > \varepsilon_{nl}$, with
$\varepsilon_{nl} =15\varepsilon_c$. The parameters are $\beta=10$, $r=0.01$, $\nu=1.19\cdot 10^{-6}$ and the forcing
is an isotropic ring in wavenumber space with radius $K_f=10$ and width $\Delta K_f = 1$.}
\label{fig:zmf}
\end{figure}

\begin{figure}
 \centering
\includegraphics[width=19pc]{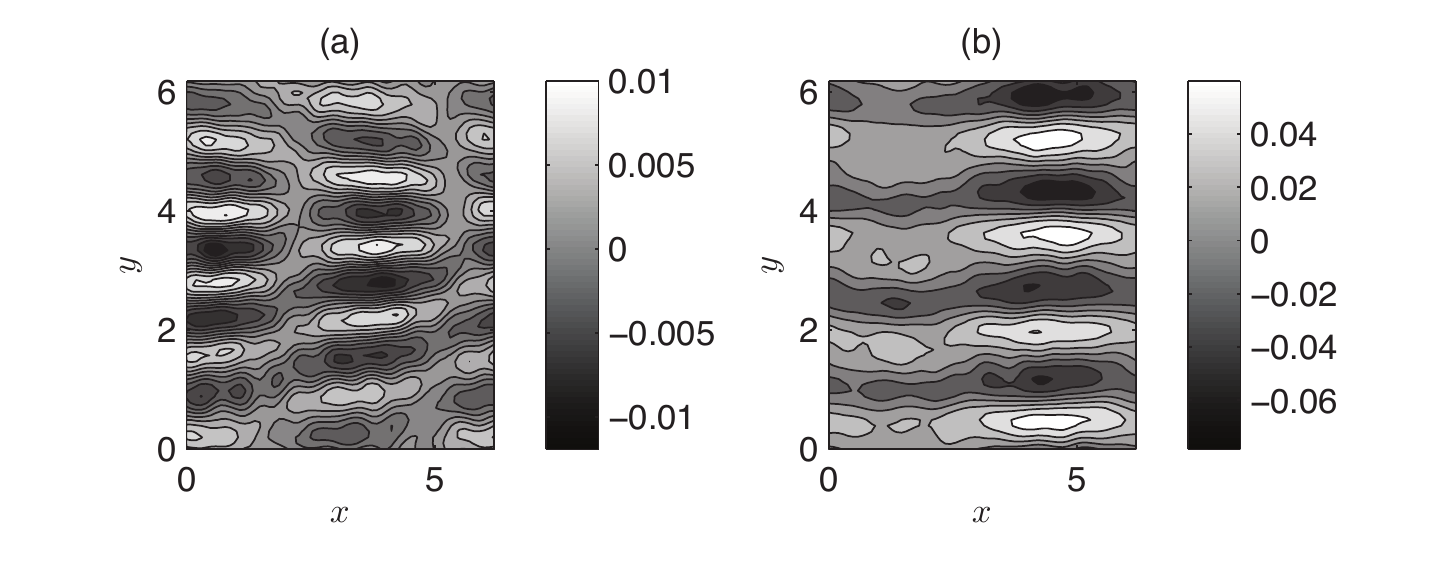}
\caption{Snapshot of the mean streamfunction $\Psi$ at statistical equilibrium obtained from the ensemble mean
quasi-linear simulations with $N_{\textrm{ens}}=10$ members for $\varepsilon=4\varepsilon_c$ (panel a) and
$\varepsilon=50\varepsilon_c$ (panel b). The parameters are as in figure~\ref{fig:zmf}.}
\label{fig:emql_struc}
\end{figure}

We choose the same parameter values as in the S3T integrations in section 5
($\beta=10$, $r=0.01$, $\nu=1.19\cdot10^{-6}$, $K_f=10$ and $\Delta K_f=1$). For these parameters
($\tilde\beta=100$), S3T predicts the emergence of propagating non-zonal structures when the energy input rate
exceeds the critical threshold $\tilde\varepsilon_c$, and the emergence of mixed zonal jet-traveling wave states
when the finite amplitude traveling wave states become structurally unstable to zonal jet perturbations. In
order to examine whether the same bifurcations occur in the EQL system, we consider two indices that
measure the power concentrated at scales larger than the scales forced. The first is the zonal mean flow
index defined as in \cite{Srinivasan-Young-12}, as  the ratio of the energy of zonal jets with scales
larger than the scale of the forcing over the total energy
\begin{equation}
\mbox{zmf}=\frac{\sum_{l:l<K_f-\Delta K_f} \hat{E}(k=0,l)}{\sum_{kl}\hat{E}(k,l)},\label{eq:zmf}
\end{equation}
where
\begin{equation}
\hat{E}(k, l)=\frac{1}{2T}\int_0^T\left(\left<\frac{|\hat\zeta '|^2}{k^2+l^2}\right>+\frac{|\hat Z|^2}{k^2+l^2}\right)dt
\end{equation}
is the time averaged total energy power spectrum of the flow at wavenumbers $(k,l)$. The
second is the non-zonal mean flow index defined as the ratio of the energy of the non-zonal modes with scales
larger than the scale of the forcing over the total energy:
\begin{equation}
\mbox{nzmf}=\frac{\sum_{kl: K<K_f-\Delta K_f}\hat{E}(k,l)}
{\sum_{kl}\hat{E}(k,l)}-\mbox{zmf}.\label{eq:nzmf}
\end{equation}
If the structures that emerge are coherent, then these indices quantify their amplitude. Figure \ref{fig:zmf}
shows both indices as a function of the energy input rate $\varepsilon$ and as a function of the corresponding
values of the zonostrophy index $R_\beta$ for EQL simulations with $N_{\textrm{ens}}=10$
members. The rapid increase of the nzmf index for $\varepsilon>\varepsilon_c$ (corresponding to
$R_\beta>1.55$), illustrates that this regime transition in the flow predicted by S3T with the emergence of
non-zonal structures manifests
in the quasi-linear dynamics as well. We now consider the case $\varepsilon=4\varepsilon_c$ in detail in
which the traveling wave structure $Z=\cos(x)\cos(5y)$ is maintained in the S3T integrations. We observe,
that the S3T equilibria manifest in the EQL simulations with the addition of some 'thermal noise' due to the
stochasticity of the forcing that is retained in this system. This is illustrated in figure~\ref{fig:equil1}b
showing the energy growth of the coherent structure for $N_{\textrm{ens}}=10$ and $N_{\textrm{ens}}=100$. The energy of
the coherent structure in the EQL integrations fluctuates around the values predicted by the
S3T system with the fluctuations decreasing as $1/\sqrt{N_{\textrm{ens}}}$. However, even with only 10 ensemble members
we get an estimate that is very close to the theoretical estimate of the infinite ensemble members obtained from
the S3T integration. The structure of the traveling wave equilibrium in the quasi-linear simulations shown in
figure~\ref{fig:emql_struc}(a) and its phase
speed (not shown) are also in very good agreement with the corresponding structure and phase speed obtained
from the S3T integration.

The second transition in which zonal jets emerge is more intriguing. While the homogeneous
equilibrium is structurally unstable to zonal jets when $\varepsilon_{sz}=5.2\varepsilon_c$, the
finite amplitude zonal jet equilibria are structurally unstable and the flow stays on the attractor
of the non-zonal traveling wave equilibria (cf.~figure~\ref{fig:equil2}). When
$\varepsilon>\varepsilon_{nl}$, the non-zonal traveling wave equilibria become S3T unstable
while at these parameter values the S3T system has mixed zonal jet-traveling wave
equilibria which are stable (cf.~figure~\ref{fig:equil4}). The rapid increase in the zmf index with the
concomitant rapid decrease in the nzmf index shown in figure~\ref{fig:zmf}, illustrates that this
regime transition manifests in the EQL system as well with similar mixed zonal-traveling wave states
appearing. The structure of the mixed zonal jet-traveling wave equilibrium for $\varepsilon=50\varepsilon_c$
is shown in \ref{fig:emql_struc}(b) and similar to the S3T equilibrium in figure~\ref{fig:equil4},
it consists mainly of 4 zonal jets and the compact support vortices $Z\sim \cos(x)\cos(4y)$ embedded in
the jets. We therefore conclude that the EQL system accurately captures the characteristics of the emerging
structures.

\subsection{Comparison to non-linear simulations}

\begin{figure}
 \centering
\includegraphics[width=19pc]{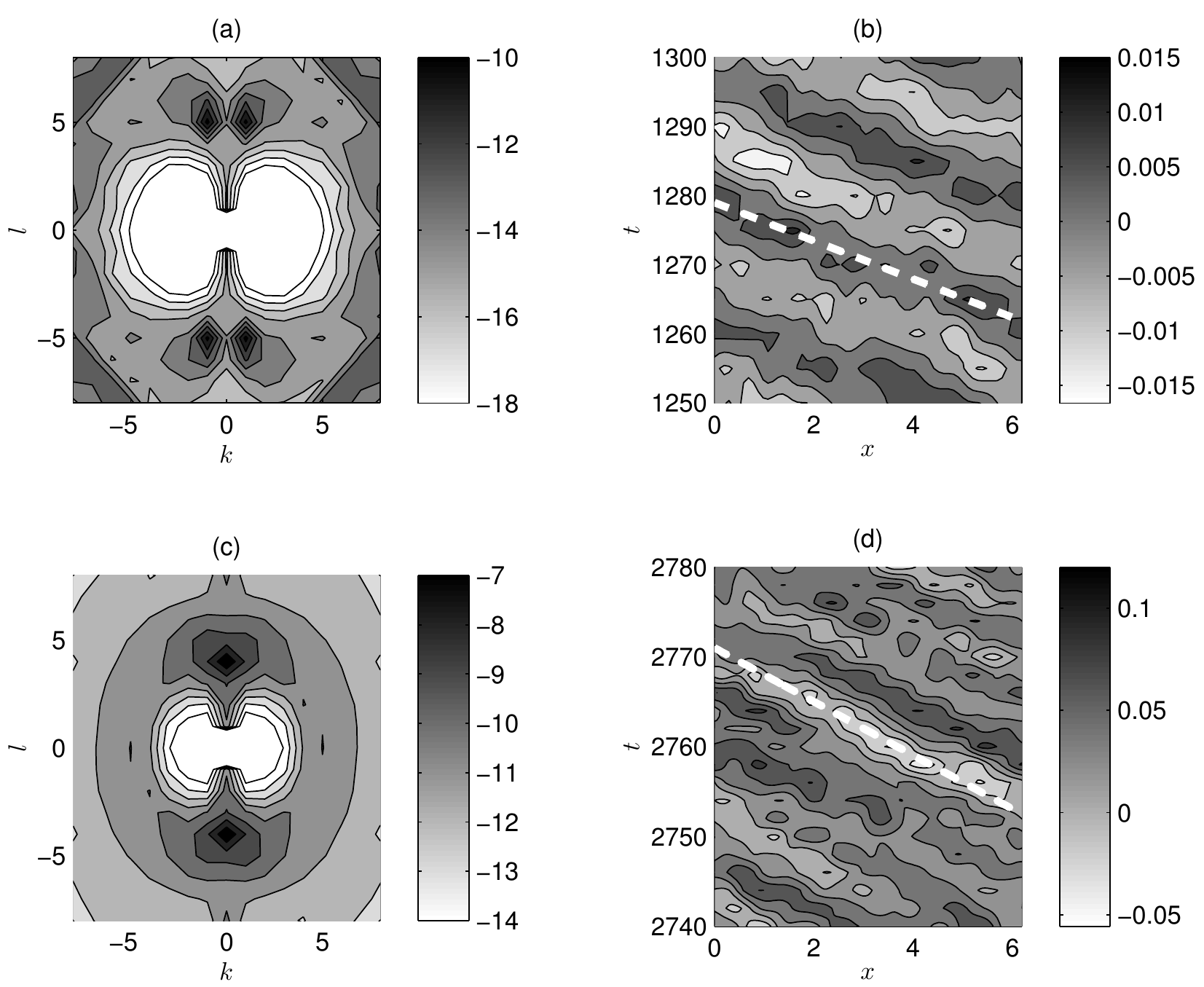}
\caption{Time averaged energy power spectra, $\log(\hat{E}(k,l))$, obtained from the non-linear
(NL) simulation of (\ref{eq:derivation1}) at $\varepsilon/ \varepsilon_c=4$ (panel a) and
$\varepsilon/ \varepsilon_c=50$ (panel c). Hovm\"oller diagram of $\psi(x,y=\pi/4,t)$ at
$\varepsilon/ \varepsilon_c=4$ (panel b) and $\varepsilon/ \varepsilon_c=50$ (panel d). The thick dashed
lines correspond to the phase speed obtained from the eigenvalue relation (\ref{eq:dispersion}).}
\label{fig:NL_snap1}
\end{figure}

In order to compare the predictions of S3T to the non-linear simulations, we solve (\ref{eq:derivation1})
with the narrow band forcing (\ref{eq:finite_ring}) on a doubly periodic channel of size $2 \pi \times 2 \pi$
using the same pseudospectral code as in the EQL simulations and the same parameter values. Figure \ref{fig:zmf}
shows the nzmf and zmf indices as a
function of the energy input rate $\varepsilon$ for the NL simulations. The rapid increase in the
nzmf index for $\varepsilon>\varepsilon_c$ shows that the non-linear dynamics share the same bifurcation
structure as the S3T statistical dynamics. In addition, the stable S3T equilibria are in principle viable
repositories of energy in the turbulent flow and the non-linear system is expected to visit their attractors for
finite time intervals. Indeed for $\varepsilon=4\varepsilon_c$, the pronounced peak at $(|k|, |l|)=(1, 5)$ of the
time averaged power spectrum shown in figure~\ref{fig:NL_snap1}(a) illustrates that the traveling wave equilibrium
with $(|k|, |l|)=(1, 5)$ that emerges in the S3T integrations, is the dominant structure in the NL simulations.
Comparison of the energy spectra obtained from the EQL and the NL simulations (not shown), reveals that the amplitude
of this structure in the quasi-linear and in the non-linear dynamics almost matches. Remarkably, the phase speed of
the S3T traveling wave matches with the corresponding phase
speed of the $(|k|, |l|)=(1, 5)$ structure observed in the NL simulations, as can be seen in the
Hovm\"oller diagram in figure~\ref{fig:NL_snap1}(b). Such an agreement in the
characteristics of the emerging structures between the EQL and NL simulations occurs for a wide range of
energy input rates as can be seen by comparing the nzmf indices in figure~\ref{fig:zmf}. As a result, S3T predicts
the dominant non-zonal propagating structures in the non-linear simulations, as well as their amplitude and phase speed.

We now focus on the second regime transition with the emergence of zonal jets. The increase in the
$\mbox{zmf}$ index in the NL simulations for $\varepsilon>\varepsilon_{nl}$ that is shown in figure
\ref{fig:zmf}, indicates the emergence of jets roughly at the bifurcation point
of the S3T and EQL simulations. However, the energy input rate threshold for the emergence of jets is
larger in the NL simulations compared to the corresponding EQL threshold. This discrepancy
possibly occurs due to the fact that the exchange of instabilities between the
mixed jet-traveling wave equilibria and the pure traveling wave equilibria depends on the equilibrium structure
$[Z^E, C^E]$. Small changes for example in $C^E$ that might be caused by the eddy-eddy terms neglected in
S3T can cause the exchange of instabilities to occur at slightly different energy input rates. It was shown in a
recent study that when the effect of the eddy-eddy terms is taken into account by obtaining $C^E$ directly from the
nonlinear simulations, the S3T stability analysis performed on this corrected equilibrium states accurately predicts
the energy input rate for the emergence of jets in the nonlinear simulations \citep{Constantinou-etal-2012}.
The power spectrum obtained from the NL simulations for $\varepsilon=50\varepsilon_c$ shows an energy peak
at $(k, |l|)=(0, 4)$ with secondary power peaks at $(|k|,|l|)=(1, 4)$ and $(|k|,|l|)=(1, 5)$) (of
approximately $12\%$ of the energy in the zonal jet each). The Hovm\"oller diagram of the streamfunction shown
in figure~\ref{fig:NL_snap1}(d) reveals that the dominant non-zonal structures in the NL simulations propagate
in the retrograde direction. As a result the mixed S3T equilibrium of figure~\ref{fig:equil4} manifests in the NL
simulations. Note however, that the phase speed calculated from the diagram is different than the phase
speed of the $(|k|, |l|)=(1, 4)$ structure in figure~\ref{fig:equil4}. At larger
energy input rates the zonal jets have typically larger scales due to jet merging and coexist with energetically
significant westward propagating non-zonal structures having an energy between $10-50 \%$ of the jet energy and
scales $(|k|, |l|)=(1, m)$, where $m$ is the number of jets in the channel. Such an agreement again holds for
a wide range of energy input rates, as the zmf indices obtained from the EQL and the NL simulations indicate.
In summary, S3T predicts the characteristics of both non-zonal propagating structures and of zonal jets in the
non-linear simulations.

\subsection{Zonostrophic regime}

\begin{figure}
 \centering
\includegraphics[width=19pc]{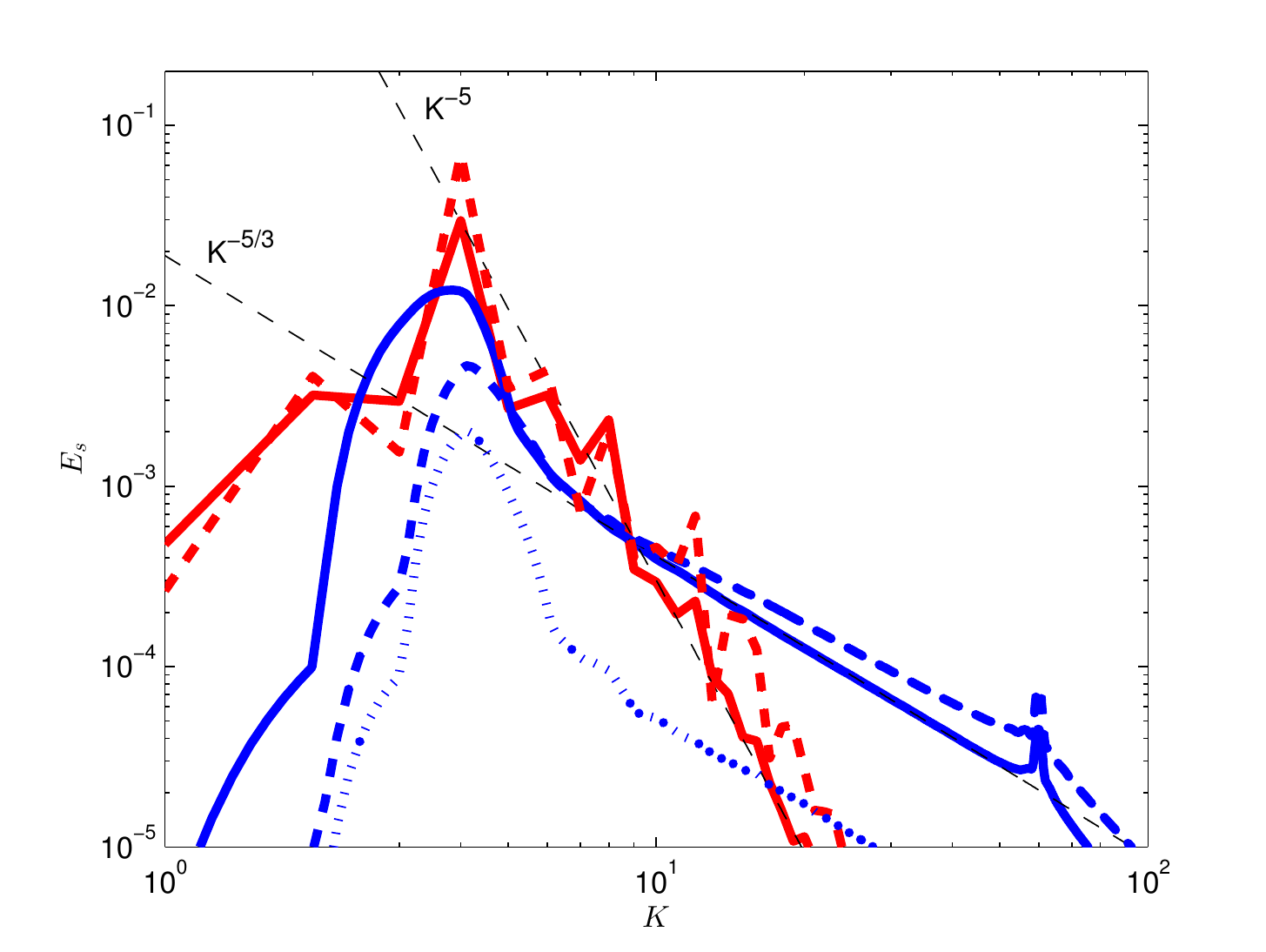}
\caption{Residual (blue) and zonal (red) energy spectra for the NL (solid) and EQL (dashed) simulations in the zonostrophic
regime. Also shown are the residual spectra from the EQL simulations when only the coherent motions are taken into account
(blue dotted line). The parameters are $K_f=60$, $\Delta K=1$, $\beta=42$, $r=0.01$, $\varepsilon=0.0065$, for which
$\tilde\beta=70$, $\tilde\varepsilon=7308\tilde\varepsilon_c$, $k_\beta=12.9$ and $R_\beta=2.5$. Lines
(thin dashed) of slope $K^{-5/3}$ and $K^{-5}$ are also plotted for reference. The pseudo-spectral code was run
with a $512\times 512$ resolution and the exponential filter of \cite{Smith-etal-02} instead of hyper-diffusion.}
\label{fig:zonos}
\end{figure}

S3T and the corresponding ensemble quasi-linear system were obtained by ignoring the eddy-eddy non-linear
interactions. Therefore the question arises on whether the predictions of S3T are useful in the zonostrophic regime.
In this regime,
which is highly supercritical with respect to S3T instability of the homogeneous equilibrium (cf.~figure~\ref{fig:Rbeta}),
maintenance of zonal jets and zonons were interpreted by previous studies to arise from an inverse energy cascade \citep{Galperin-etal-10},
a highly non-linear process, which is absent in S3T. According to this interpretation, the turbulent energy cascades isotropically toward large
scales until it reaches $k_\beta$. At this scale the cascade becomes anisotropic and most of the energy is channelled into
the zonal flows. To illustrate this, the time averaged energy power spectra
$\hat{E}(k, l)$ are typically split between the zonal spectra $\hat{E}_z(l)=\hat{E}(k=0,l)$ and the residual
$\hat{E}_R(k, l)=\hat{E}-\hat{E}_z$. The zonal and residual spectra calculated from NL integrations in the zonostrophic regime
($K_f=60$, $\Delta K=1$, $\beta=42$, $r=0.01$, $\varepsilon=0.0065$) are shown
in figure~\ref{fig:zonos}. Up to the scale $k_\beta$, the residual spectra follow the
Kolmogorov $K^{-5/3}$ law in accordance with an isotropic cascade assumption. At this scale, the cascade is
anisotropized and the residual spectra steepen. However, most of the energy is in zonal scales with the zonal spectra
following a much steeper $K^{-5}$ law.

The residual and the zonal spectra obtained from an EQL simulation with $N_{\textrm{ens}}=10$ for the same parameters, are also
shown in figure~\ref{fig:zonos}. The residual spectra follow a shallower than $K^{-5/3}$ slope for $K>k_\beta$, while
they steepen after $k_\beta$ and reach a lower peak with respect to the corresponding
spectra from the NL simulations. In addition, the residual part of the spectrum corresponds mainly to incoherent
motions for scales with $K>k_\beta$. This is revealed by taking into account only the spectra of the coherent part of
the flow and calculating the residual spectrum that is also shown in figure~\ref{fig:zonos}. For most of the scales, it
is at least one to two orders of magnitude lower than the corresponding residual spectrum when both coherent and
incoherent motions are taken into account and only the non-zonal structures with large scales (close to the energy peak)
appear to be coherent. Failure of the EQL simulations to exactly reproduce the $K^{-5/3}$ slope of the incoherent
turbulent motions is not surprising, since  the inverse energy cascade that is absent in the EQL
simulations is essential for this part of the spectrum. The energetically important part however, which contains the
large-scale energetic waves is captured by S3T. The zonal spectra obtained from the EQL simulations follow the same
$K^{-5}$ law and peak at the same scale compared to the NL simulations but the peak has a larger amplitude. As
is argued in section 5.2.2, the steep power law is an artifact of the shape of the strongly forced jet which
is characterized by near discontinuity in the shear at the maxima of the prograde jets.

So to summarize, the scale
and the shape of the dominant jet structure, as well as the scale of the most energetic
coherent non-zonal structures are accurately captured by the EQL simulations, while the eddy-eddy interactions
neglected in the EQL simulations set the proper scaling for the tail of the spectrum that consists of incoherent
turbulent motions and change the partition between the energy of the jet (that is overestimated in the EQL simulations) and
that of the non-zonal large-scale structures (that is underestimated in the EQL simulations).

\section{Summary -- Discussion}

This article addressed the emergence of coherent structures in barotropic $\beta$-plane
turbulence using the tools of Stochastic Structural Stability Theory (S3T), a statistical
theory that expresses the statistics of the turbulent flow dynamics as a systematic cumulant
expansion truncated at second order. With the interpretation of the ensemble average as a Reynolds average
over the fast turbulent eddies adopted in this contribution, the second order cumulant expansion results
in a closed, non-linear dynamical system that governs the joint evolution of slowly varying, spatially
localized coherent structures with the second order statistics of the rapidly evolving turbulent eddies.
The fixed points of this autonomous, deterministic non-linear system define statistical equilibria,
the stability of which are amenable to the usual treatment of linear and non-linear stability analysis.

The linear stability of the homogeneous S3T equilibrium with no mean velocity was examined analytically
for the case of an isotropic random stirring that sustains the turbulence in the barotropic flow.
Structural instability was found to occur for perturbations with smaller scale
than the forcing, when the energy input rate $\tilde\varepsilon=\varepsilon K_f^2/r^3$ is larger than a certain threshold
$\tilde\varepsilon_c$ that depends on $\tilde\beta=\beta /(rK_f)$. It was found that when $\tilde\beta$ is small or order
one, the maximum growth rate occurs for stationary zonal structures, while for large $\tilde\beta$, westward
propagating non-zonal grow the most.

The eddy--mean flow dynamics underlying the S3T instability of zonal jets was then studied in detail. It was shown that
close to the structural stability boundary, the dynamics can be split into two competing processes. The
first, which is shearing of the eddies by the local shear described by Orr dynamics in the $\beta$ plane, was shown
to lead to jet forming upgradient momentum fluxes acting exactly as negative viscosity for an anisotropic forcing and
as negative hyperviscosity for isotropic forcing. The second, which is momentum flux divergence resulting
from lateral wave propagation on the nonuniform local mean vorticity gradient, was shown to lead to jet opposing
downgradient fluxes acting as hyperdiffusion.

The equilibration of the unstable, exponentially
growing coherent structures for large $\tilde\beta$ was then studied through numerical integrations of the S3T dynamical
system. When the forcing amplitude is slightly
supercritical, the finite amplitude traveling wave equilibrium has a structure close
to the corresponding unstable non-zonal perturbation with the same scale. When the forcing amplitude is highly
supercritical, the instabilities equilibrate to mixed states consisting of strong zonal jets with
smaller amplitude traveling waves embedded in them.

The predictions of S3T were then compared to the results obtained from direct numerical simulations of the turbulent
dynamics. The critical
threshold above which coherent non-zonal structures are unstable according to the stability analysis of the
S3T system was found to be in excellent agreement with the critical value above which non-zonal structures acquire
significant power in the non-linear simulations. The scale, phase speed and amplitude of the dominant structures
in the non-linear simulations were also found to correspond to the structures predicted by S3T. In addition, the threshold for
the emergence of jets, which is identified in S3T as the energy input rate at which an S3T stable, finite amplitude
zonal jet equilibrium exists, was found to roughly match the corresponding threshold for jet formation in the non-linear
simulations, with the emerging jet scale and amplitude being accurately obtained using S3T. Such a good agreement
between the predictions of S3T and direct numerical simulations, holds not only close to the bifurcation point
for the emergence of coherent structures but also in the regime of zonostrophic
turbulence. Consequently, these results provide a concrete example that large-scale structure in barotropic turbulence,
whether it is zonal jets or non-zonal coherent structures, emerges and is sustained from systematic self-organization
of the turbulent Reynolds stresses by spectrally non-local interactions and in the absence of a turbulent cascade.


\ifthenelse{\boolean{dc}}
{}
{\clearpage}

\begin{appendix}[A]
\addcontentsline{toc}{section}{\protect\numberline{}Appendices}
\section*{\begin{center}Boundedness of the solutions and invariants of the S3T equations\end{center}\label{appA}}

The S3T system in the absence of forcing and dissipation has similar quadratic invariants
with the nonlinear system. Further, the solutions of the S3T equations remain bounded for all times.  That is, the  sum
of the enstrophy of the ensemble mean over the domain, $H_m=1/2 \int  {Z^2} dxdy$, and the eddy
enstrophy over the domain, $H_p=1/2 \int  {C_{\mathbf{x}_1=\mathbf{x}_2}} dxdy$, is conserved.
Similarly, the sum of the energy of the ensemble mean, $E_m=1/2 \int {(U^2+V^2 )} dxdy$,
and the eddy energy, $E_p=1/2 \int ( \Delta_1^{-1} {C)_{\mathbf{x}_1=\mathbf{x}_2}} dxdy$,
is also conserved. We show this by first multiplying
(\ref{eq:Q_evo2}) (in the
absence of hyper-diffusion) by $Z$ to obtain:
\begin{equation}
\partial_t\eta_m+U\partial_x\eta_m+V\partial_y\eta_m+\beta VZ=-Z\nabla\cdot\left<{\bf u}'\zeta'\right>-
2r\eta_m,\label{eq:Q_evo_app}
\end{equation}
where $\eta_m=Z^2/2$ is the enstrophy density of the ensemble mean. Integrating by parts and using the
continuity equation we rewrite the advection terms as:
\begin{equation}
U\partial_x\eta_m+V\partial_y\eta_m=\partial_x(U\eta_m)+\partial_y(V\eta_m).
\end{equation}
Writing $Z=\partial_xV-\partial_yU$ and using again the continuity equation we have:
\begin{equation}
 ZV
=  \partial_x\frac{U^2+V^2}{2}-
\partial_y(UV) ,
\end{equation}
and  (\ref{eq:Q_evo_app}) becomes:
\begin{equation}
\partial_t\eta_m+\nabla\cdot({\bf U}\eta_m)+\beta\partial_xe_m-\beta\partial_y(UV)=
-Z\nabla\cdot\left<{\bf u}'\zeta'\right>-2r\eta_m,\label{eq:Q_evo_app2}
\end{equation}
where $e_m=(1/2)(U^2+V^2)$ is the energy density of the ensemble mean. Similarly it can be shown
from (\ref{eq:cov_evo2}), that the ensemble mean of the eddy enstrophy density
$\eta_p=(1/2)C_{\mathbf{x}_1=\mathbf{x}_2}$ evolves (in the absence of hyper-diffusion) according to:
\begin{align}
 &\partial_t\eta_p+\nabla\cdot({\bf U}\eta_p)+\beta\left[\partial_x(e_p)-\partial_y\left<u'v'\right>\right]
+\nonumber\\
  &\qquad+\left<u'\zeta '\right>\partial_xZ+
\left<v'\zeta '\right>\partial_yZ=\eta_f -2r\eta_p,\label{eq:q_evo2}
\end{align}
where $e_p=(1/2)(\Delta_1^{-1}C)_{\mathbf{x}_1=\mathbf{x}_2}$ is the ensemble mean of the eddy energy density
and $\eta_f=(1/2)\Xi_{\mathbf{x}_1=\mathbf{x}_2}$ is the enstrophy density of the forcing. Adding (\ref{eq:Q_evo_app2}) and
(\ref{eq:q_evo2}) we obtain the equation for the evolution of the total enstrophy density $\eta=\eta_p+\eta_m$:
\begin{equation}
(\partial_t+2r)\eta-\eta_f=-\nabla\cdot({\bf U}\eta)-\beta\partial_x(e_p+e_m)+
\beta\partial_y(UV+\left<u'v'\right>).\label{eq:ens}
\end{equation}
Integrating (\ref{eq:ens}) over the horizontal domain, the terms on the rhs of (\ref{eq:ens}) integrate to
zero and the total enstrophy
$H=H_m+H_p=\int ( \eta_m + \eta_p ) dxdy$
evolves according to:
\begin{equation}
\partial_tH=H_f-2rH,\label{eq:tot_ens}
\end{equation}
where $H_f$ is the total enstrophy imparted by the forcing. As a result, the enstrophy is bounded
and has the value $H^{eq}=H_f/(2r)$ at steady state. Similarly, it can be shown that the total
energy $E=E_m+E_p$ is bounded.

\end{appendix}

\begin{appendix}[B]
\section*{\begin{center}Calculation of the dispersion relation and its properties\end{center}\label{appB}}

In this Appendix we derive the dispersion relation (\ref{eq:dispersion}), which determines the stability of
zonal as well as non-zonal perturbations in homogeneous turbulence.
We follow closely the treatment of
\cite{Srinivasan-Young-12}. We first rewrite (\ref{eq:Q_evo2}), (\ref{eq:cov_evo2}) in terms of the variables
$\tilde{x}=x_1-x_2$, $\overline{x}=(1/2)(x_1+x_2)$, $\tilde{y}=y_1-y_2$ and $\overline{y}=(1/2)(y_1+y_2)$. The
derivatives transform into this new system of coordinates to $\partial_{x_i}=(1/2)\partial_{\overline{x}}+
(-1)^{i+1}\partial_{\tilde{x}}$, $\partial_{y_i}=(1/2)\partial_{\overline{y}}+(-1)^{i+1}\partial_{\tilde{y}}$, $\Delta_i=\tilde{\Delta}+(1/4)\overline{\Delta}+(-1)^{i+1}\partial_{\tilde{y}\overline{y}}^2+
(-1)^{i+1}\partial_{\tilde{x}\overline{x}}^2$, with $\tilde{\Delta}=\partial_{\tilde{x}\tilde{x}}^2+
\partial_{\tilde{y}\tilde{y}}^2$ and $\overline{\Delta}=\partial_{\overline{x}\overline{x}}^2+
\partial_{\overline{y}\overline{y}}^2$. It
is also convenient to introduce the streamfunction covariance
$S(\tilde{x}, \overline{x}, \tilde{y}, \overline{y})\equiv\left<\psi_1'\psi_2'\right>$, which is related to
$C(\tilde{x}, \overline{x}, \tilde{y}, \overline{y})$ via:
\begin{align}
C&= \left<\zeta_1'\zeta_2'\right>=\left<\Delta_1\psi_1'\Delta_2\psi_2'\right>=
\Delta_1\Delta_2 S\nonumber\\
 & =\left[\left({\tilde{\Delta}}+{1\over 4}\overline{\Delta}\right)^2-\Gamma^2\right] S,
\end{align}
where $\Gamma=\partial_{\tilde{x}\overline{x}}^2+\partial_{\tilde{y}\overline{y}}^2$.
Equations (\ref{eq:Q_evo2}), (\ref{eq:cov_evo2}) then become in the absence of hyper-viscosity ($\nu=0$):
\begin{align}
& \left[\partial_t+\overline{U}\partial_{\overline{x}}+\tilde{U}\partial_{\tilde{x}}+\overline{V}
\partial_{\overline{y}}+\tilde{V}\partial_{\tilde{y}}\right]C+\nonumber\\
 &\quad+\left[(\beta+\overline{Z}_y)\partial_{\overline{x}}+
\tilde{Z}_y\partial_{\tilde{x}}-\overline{Z}_x\partial_{\overline{y}}-\tilde{Z}_x\partial_{\tilde{y}}\right]
\left(\tilde{\Delta}+\frac{1}{4}\overline{\Delta}\right)S-\nonumber\\
 &\quad-\left[2(\beta+\overline{Z}_y)\partial_{\tilde{x}}+
\frac{1}{2}\tilde{Z}_y\partial_{\overline{x}}-2\overline{Z}_x\partial_{\tilde{y}}-\frac{1}{2}
\tilde{Z}_x\partial_{\overline{y}}\right]
\Gamma S =\nonumber\\
 &\quad=-2rC+\Xi,
\label{eq:dcdt_y}
\end{align}
\begin{equation}
\left(\partial_t+{\bf U}\cdot\nabla\right)Z+\beta V=(\partial_{\tilde{x}\overline{y}}^2-\partial_{\tilde{y}\overline{x}}^2)\Gamma S|_{\tilde{x}=\tilde{y}=0}-rZ,\label{eq:dudt_y}
\end{equation}
where $(\overline{U}, \overline{V})=(1/2)(U_1+U_2, V_1+V_2)$, $(\tilde{U}, \tilde{V})=(U_1-U_2, V_1-V_2)$,
$(\overline{Z}_x, \overline{Z}_y)=(1/2)(\partial_{x_1}+\partial_{x_2}, \partial_{y_1}+\partial_{y_2})Z$ and
$(\tilde{Z}_x, \tilde{Z}_y)=(\partial_{x_1}-\partial_{x_2}, \partial_{y_1}-\partial_{y_2})Z$.

The forcing covariance $\Xi$ is homogeneous and as a result it depends only on the difference coordinates, $\tilde x$ and $\tilde y$. It
can then be readily shown from (\ref{eq:dcdt_y})-(\ref{eq:dudt_y}), that the state with no coherent flow ($U^E=V^E=Z^E=0$) and with the
homogeneous vorticity covariance $C^E(\tilde{x}, \tilde{y})=\Xi /(2r)$ (implying also that the streamfunction covariance $S^E$ is homogenous)
is a fixed point of the S3T system. The stability of this homogeneous equilibrium, can be addressed by first linearizing the
S3T system about it:
\begin{align}
\partial_t\delta C&=-\left(\delta \tilde{U}\partial_{\tilde{x}}+\delta\tilde{V}\partial_{\tilde{y}}
\right)C^E-\left(\delta\tilde{Z}_y\partial_{\tilde{x}}-\delta \tilde{Z}_x
\partial_{\tilde{y}}\right)\tilde{\Delta}S^E-\nonumber\\
 &\qquad-\beta \left\{\left[\tilde{\Delta}+\frac{1}{4}\overline{\Delta}\right]\partial_{\overline{x}}-
2\Gamma\partial_{\tilde{x}}
\right\}\delta S-2r\delta C,\label{eq:Lc}\\
\partial_t\delta Z&=-\beta \delta V+(\partial_{\tilde{x}\overline{y}}^2-\partial_{\tilde{y}\overline{x}}^2)
\Gamma\delta S|_{\tilde{x}=\tilde{y}=0}
-r\delta Z,\label{eq:Lu}
\end{align}
where $\delta Z$, $\delta\tilde{U}$, $\delta\tilde{V}$, $\delta\tilde{Z}_x$, $\delta\tilde{Z}_y$, $\delta C$ and
$\delta S$ are small perturbations in the ensemble mean vorticity, velocities and vorticity gradients and in the eddy vorticity and
streamfunction covariances respectively, and then performing an eigenanalysis of the linearized
equations (\ref{eq:Lc})-(\ref{eq:Lu}).

We consider a harmonic vorticity perturbation of the form $\delta Z=e^{inx+imy}
e^{\sigma t}$, for which:
\begin{align}
& [\delta \tilde{U}, \delta\tilde{V},\delta\tilde{Z}_x, \delta\tilde{Z}_y]=\nonumber\\
&~~=-2\left[\frac{m}{N^2},-\frac{n}{N^2},n,m\right]
\sin\left(\frac{n\tilde{x}}{2}+\frac{m\tilde{y}}{2}\right)
e^{in\overline{x}+im\overline{y}}e^{\sigma t},\label{eq:tildeU}
\end{align}
with $N^2=n^2+m^2$. Taking the same form for the streamfunction covariance perturbation 
$\delta S = S_{nm}(\tilde{x}, \tilde{y}) e^{in\overline{x}+im\overline{y}}e^{\sigma t}$ and inserting it in
(\ref{eq:Lc})-(\ref{eq:Lu}) along with
(\ref{eq:tildeU}) yields:
\begin{align}
& (\sigma+2r)\left[\left(\tilde{\Delta}-\frac{N^2}{4}\right)^2+\Delta_+^2\right]S_{nm}\nonumber\\
 &-\left[2i\beta\Delta_+
\partial_{\tilde{x}}-in\beta
\left(\tilde{\Delta}-\frac{N^2}{4}\right)\right] S_{nm} = \nonumber\\
 &~=\frac{2}{N^2}\sin\left(\frac{n\tilde{x}}{2}+\frac{m
\tilde{y}}{2}\right)\left(m\partial_{\tilde{x}}-n\partial_{\tilde{y}}\right)(\tilde{\Delta}+N^2)
\tilde{\Delta}S^E,\label{eq:disp1}\\
&-(\sigma +r)N^2+in\beta=N^2\left(m\partial_{\tilde{x}}-n\partial_{\tilde{y}}\right)\Delta_+S_{nm}
|_{\tilde{x}=\tilde{y}=0},\label{eq:disp2}
\end{align}
where $\Delta_+=n\partial_{\tilde{x}}+m
\partial_{\tilde{y}}$ and $C^E=\Xi/2r=\tilde{\Delta}^2 S^E$ is the
equilibrium vorticity covariance with zero mean flow.

Defining the Fourier transform of ${S}_{nm}(\tilde{x},\tilde{y})$ by
\begin{equation}
\hat{S}_{nm} (k,l)={1\over 2\pi} \int_{-\infty}^{\infty}\int_{-\infty}^{\infty} { S}_{nm}(\tilde{x},\tilde{y})
e^{-ik\tilde{x}-il\tilde{y}} \mathrm{d} \tilde{x} \mathrm{d} \tilde{y}~,
\end{equation}
we obtain from  (\ref{eq:disp1})  that the Fourier component $\hat{S}_{nm}$ satisfies:
\begin{align}
\hat{S}_{nm}&=\frac{(mk_--nl_-)
K_-^2(K_-^2/N^2-1)\hat{S}^E(k_-,l_-)}{i\beta (k_-K_+^2-k_+K_-^2)+
(\sigma+2r) K_+^2K_-^2}-\nonumber\\ 
& \qquad- \frac{(mk_+-nl_+)K_+^2(K_+^2/N^2-1)\hat{S}^E(k_+,l_+)}
{i\beta (k_-K_+^2-k_+K_-^2)+(\sigma+2r) K_+^2K_-^2},\label{eq:Psin}
\end{align}
with $k_\pm=k\pm n/2$, $l_\pm=l\pm m/2$, $K_\pm^2=k_\pm^2+l_\pm^2$ and
$K^2=k^2+l^2$. $\hat{S}^E=\hat{\Xi}/(2rK^4)$ is the Fourier transform
of $ S^E$, and $\hat{\Xi}$ is the Fourier transform of  $\Xi$. In addition, (\ref{eq:disp2}) becomes:

\begin{align}
& in\beta-(\sigma+r)N^2=\nonumber\\ 
&~~=-{N^2\over 2\pi}\int_{-\infty}^\infty\int_{-\infty}^\infty \left[nm(k^2-l^2)+(m^2-n^2)kl\right]
\hat{S}_{nm}dkdl\nonumber\\
 &~~=\Lambda_+-\Lambda_-,\label{eq:disper1}
\end{align}
where
\begin{align}
 \Lambda_\pm&={1\over 2\pi}\int_{-\infty}^\infty\int_{-\infty}^\infty
dkdl K_\pm^2(K_\pm^2-N^2)
\hat{S}^E(k_\pm,l_\pm)\times \nonumber\\ 
& \quad\times\frac{\left[nm(k^2-l^2)+(m^2-n^2)kl\right](mk_\pm-nl_\pm)}{i\beta(k_-K_+^2-k_+K_-^2)+(\sigma+2r)
K_+^2K_-^2}.\label{eq:lambda}
\end{align}

Equation (\ref{eq:disper1}) can be further simplified by noting that because the choice of $\mathbf{x}_1$ and
$\mathbf{x}_2$ is arbitrary, the forcing covariance satisfies the exchange symmetry $\Xi(x_1,x_2,y_1,y_2)=
\Xi(x_2,x_1,y_2,y_1)$. In terms of the new variables, the exchange symmetry is written as
$\Xi(\tilde{x}, \overline{x}, \tilde{y},\overline{y})=\Xi(-\tilde{x}, \overline{x},-\tilde{y},\overline{y})$, and
consequently $\hat{\Xi}$ satisfies $\hat{\Xi}(-k,-l)=\hat{\Xi}(k,l)$. As a result:
\begin{equation}
\Lambda_+=-\Lambda_-.\label{eq:lambda2}
\end{equation}
Using (\ref{eq:lambda2}) and shifting the axes in the resulting integrals ($k\rightarrow k+n/2$ and
$l\rightarrow l+m/2$), reduces (\ref{eq:disper1}) to the following dispersion relation:
\begin{align}
& \int_{-\infty}^{\infty}\int_{-\infty}^{\infty}dkdl \; K^2(K^2-N^2)\hat{S}^E(k,l)\times \nonumber\\
& ~\times\frac{(mk-nl)\left[nm(k_+^2-l_+^2)+(m^2-n^2)k_+l_+
\right]}{i\beta\left(kK_s^2-(k+n)K^2\right)+(\sigma+2r)K^2
K_s^2}=\nonumber\\
 &~~\qquad=\pi(\sigma+r)N^2-i\pi n\beta,\label{eq:dispersion_app}
\end{align}
where $K_s^2=(k+n)^2+(l+m)^2$. The corresponding dispersion relation on a periodic box, can be readily calculated by
simply substituting the integrals in (\ref{eq:dispersion_app}) by finite sums of integer wavenumbers. For a mirror symmetric
forcing obeying:
\begin{equation}
\hat\Xi(-k, l)=\hat\Xi(k, l),\label{eq:mirror}
\end{equation}
the eigenvalues $\sigma$ satisfy the symmetries (\ref{eq:symmet}). In
order to show this, we consider (\ref{eq:dispersion_app}) for $\sigma_{(-n, m)}$ and change the sign of $k$ in
the integral to obtain:
\begin{align}
& \int_{-\infty}^{\infty}\int_{-\infty}^{\infty}dkdl \;K^2(K^2-N^2)\hat{S}^E(-k,l)\times\nonumber\\
  & ~~\times\frac{(mk-nl)\left[nm(k_+^2-l_+^2)+(m^2-n^2)k_+l_+
\right]}{-i\beta\left(kK_s^2-(k+n)K^2\right)+(\sigma_{(-n, m)}+2r)K^2
K_s^2}=\nonumber\\ 
&~~\qquad=\pi(\sigma_{(-n, m)}+r)N^2+i\pi n\beta.\label{eq:symm_app1}
\end{align}
Taking the conjugate of (\ref{eq:symm_app1}) and using the mirror symmetry (\ref{eq:mirror}) yields
(\ref{eq:dispersion_app}) and therefore $\sigma_{(-n, m)}=\sigma_{(n, m)}^*$. Similarly, it can be readily
shown by considering (\ref{eq:dispersion_app}) for $\sigma_{(n, -m)}$ and changing the sign of $l$ in
the integral, that $\sigma_{(n, -m)}=\sigma_{(n, m)}$.

\end{appendix}


%


\end{document}